\documentclass[11pt,a4paper]{article}
\usepackage{etex}
\usepackage[thinlines]{easybmat}
\usepackage[T1]{fontenc}
\usepackage{pstricks-add}
\usepackage[body={16cm,23cm}]{geometry}
\usepackage{enumerate}
\usepackage{amsmath,amssymb}
\usepackage{bm}
\usepackage{epic}
\usepackage{graphicx}
\usepackage{color}
\usepackage{hyperref}
\usepackage{fixfoot}

\newcommand{\specbaz}{z}

\newcommand{\Qop}{{\rm \bf{Q}}}

\newcommand{\Top}{{\rm {\bf T}}}

\newcommand{\Lbf}{{\mathbf L}}

\newcommand{\Dbf}{{\mathcal D}}
\newcommand{\Rbf}{{\mathbf R}}

\newcommand{\Xbf}{{\mathbf X}}
\newcommand{\beq}{\begin{equation}}
\newcommand{\eeq}{\end{equation}}

\newcommand{\Qf}{ {\bf Q}}

\newcommand{\nstr}{{\rm \widehat{Str}}}

\newcommand{\alg}[1]{\mathfrak{#1}}

\newcommand{\osca}{\mathbf{a}}

\newcommand{\oscc}{\mathbf{c}}

\newcommand{\osch}{\mathbf{h}}
\newcommand{\suposc}{\mathbf{\xi}}

\newcommand{\sfrac}[2]{{\textstyle\frac{#1}{#2}}}
\newcommand{\half}{\sfrac{1}{2}}

\newcommand{\p}{{p}}

\DeclareFixedFootnote{\repfoot}{$\osca\,\osca^\dagger-\osca^\dagger\,\osca=1$ and $\oscc\,\oscc^\dagger+\oscc^\dagger\,\oscc=1$.}

\newcommand{\bea}{\begin{eqnarray}}
\newcommand{\eea}{\end{eqnarray}}

\makeatletter
\newcommand{\raisemath}[1]{\mathpalette{\raisem@th{#1}}}
\newcommand{\raisem@th}[3]{\raisebox{#1}{$#2#3$}}
\makeatother

\makeatletter
\def\mr@ignsp#1 {\ifx\:#1\@empty\else #1\expandafter\mr@ignsp\fi}%
\newcommand{\multiref}[1]{\begingroup
\xdef\mr@no@sparg{\expandafter\mr@ignsp#1 \: }%
\def\mr@comma{}%
\@for\mr@refs:=\mr@no@sparg\do{\mr@comma\def\mr@comma{,}\ref{\mr@refs}}%
\endgroup}
\makeatother

\numberwithin{equation}{section}

\begin{document}
\psset{arrowscale=2}
\thispagestyle{empty}
\pagenumbering{alph}

\begin{flushright}\footnotesize
\texttt{HU-Mathematik:~2010-13}\\
\texttt{HU-EP-10/42}\\
\texttt{AEI-2010-126}\\
\vspace{0.5cm}
\end{flushright}
\setcounter{footnote}{0}

\begin{center}
{\Large\textbf{\mathversion{bold}
Oscillator Construction of \texorpdfstring{$\mathfrak{su}(n|m)$}{} Q-Operators
}\par}
\vspace{15mm}

{\sc Rouven Frassek $^{a,b}$, Tomasz {\L}ukowski $^{a}$, Carlo Meneghelli $^{a,b}$,\\
Matthias Staudacher $^{a,b}$}\\[5mm]

{\it $^a$ Institut f\"ur Mathematik und Institut f\"ur Physik, Humboldt-Universit\"at zu Berlin\\
Johann von Neumann-Haus, Rudower Chaussee 25, 12489 Berlin, Germany
}\\[5mm]

{\it $^b$ Max-Planck-Institut f\"ur Gravitationsphysik, Albert-Einstein-Institut\\
    Am M\"uhlenberg 1, 14476 Potsdam, Germany}\\[5mm]

\texttt{rfrassek@physik.hu-berlin.de}\\
\texttt{lukowski@mathematik.hu-berlin.de}\\
\texttt{carlo@aei.mpg.de}\\
\texttt{matthias@aei.mpg.de}\\[58mm]

\textbf{Abstract}\\[2mm]
\end{center}

\noindent{
We generalize our recent explicit construction of the full hierarchy of Baxter Q-operators of compact spin chains with $\alg{su}(n)$ symmetry to the supersymmetric case $\alg{su}(n|m)$. The method is based on novel degenerate solutions of the graded Yang-Baxter equation, leading to an amalgam of bosonic and fermionic oscillator algebras. Our approach is fully algebraic, and leads to the exact solution of the associated compact spin chains while avoiding Bethe ansatz techniques. It furthermore elucidates the algebraic and combinatorial structures underlying the system of nested Bethe equations. Finally, our construction naturally reproduces the representation, due to Z.~Tsuboi, of the hierarchy of Baxter Q-operators in terms of hypercubic Hasse diagrams.
}

\newpage
\clearpage\pagenumbering{arabic}
\setcounter{page}{1}


\section{Introduction, Motivation, Overview, and Outlook}
\label{sec:intro}

Quantum integrability is a very rich and intricate phenomenon, which was, surely somewhat serendipitously, discovered some 80 years ago by Hans Bethe \cite{Bethe}. Its underpinnings and underlying mathematical structures continue to be unearthed, and it is fair to say that no fully general theory of quantum integrability exists to date. This is unfortunate, since integrability keeps reappearing in surprising and important contexts within theoretical physics. A prime example is the AdS/CFT correspondence. An overview of a very recent collection of up-to-date review articles on this exciting and still quite mysterious appearance is \cite{Beisert:2010jr}. In view of the sheer variety of approaches and the manifest lack of an ab initio, transparent, constructive, and self-contained solution, it should be obvious that the underlying fundamental principles of gauge/string integrability have not been discovered yet.

It so turns out, that even the theory of nearest-neighbor quantum spin chains is not yet complete. This is an important clue, as spin chains appear in the weak coupling limit of the AdS/CFT system \cite{Minahan:2010js}. In \cite{Bazhanov:2010ts} we presented an explicit construction of the two Baxter Q-operators of the $\alg{su}(2)$ Heisenberg XXX spin chain, historically the first model solved by Bethe's ansatz \cite{Bethe}. For a very elementary review, see  \cite{Staudacher:2010jz}. While Rodney Baxter introduced the notion of the Q-operator in his seminal article \cite{Baxter:1972hz} on the XYZ chain (alias the 8-vertex model, see also his textbook \cite{Baxter:book}), the limiting procedure back to the XXX chain is not straightforward at all. In fact, our construction is completely different from Baxter's orginal one. In the course of generalizing \cite{Bazhanov:2010ts} to the $\alg{su}(n)$ case in \cite{Bazhanov:2010jq}, the Q-operator construction method was put on firm ground. Four of its key features deserve special mentioning.

The {\it first feature} is that the set of $\alg{su}(n)$ Q-operators is constructed just like any ``ordinary'' transfer matrix as the trace over monodromies built from products of Lax operators ${\bf L}(z)$, just as in Baxter's work \cite{Baxter:1972hz} and in the quantum inverse scattering approach \cite{Faddeev:1996iy}, where however the Lax operators now correspond in general to novel {\it degenerate} solutions of the Yang-Baxter equation. $z$ is the spectral parameter. The fact that these new types of Lax operators were not previously known is what we had in mind when stating above that the theory of integrable quantum spin chains has not yet been completed. Bringing them this way into the standard framework \cite{Faddeev:1996iy} allows to naturally include the Baxter Q-operators into the set of all mutually commuting operators of a given spin chain. It furthermore permits the derivation of the hierarchy of functional fusion relations on the operatorial level with the help of certain {\it factorization formulas} \cite{Bazhanov:2010jq}. Our method therefore directly reproduces and {\it explains} the full set of functional equations of the $\alg{su}(n)$ spin chain models, which was previously known only on the level of eigenvalues (as opposed to operators), see \cite{Krichever:1996qd} and references therein.

The {\it second feature} is a consequence of the first feature: the new degenerate solutions of the Yang-Baxter equation lead in general on purely algebraic grounds to {\it bosonic oscillator} degrees of freedom in the auxiliary channel of the Lax operators. These oscillators initially appeared in the so-called BLZ construction of conformal field theory \cite{Bazhanov:1998dq}, and our new ``Yangian'' point of view \cite{Bazhanov:2010jq} therefore also contributes to a deeper explanation of the latter. Note that all physically sensible representations of bosonic oscillators are infinite dimensional. We find it fascinating that these representations are needed to fully describe the integrable structure of {\it compact}  $\alg{su}(n)$ spin chain models. We consider this to be an important hint on why spin chains may appear from quantum sigma models, where infinite dimensional representations are required as in any other quantum field theory. This is precisely what happens in the AdS/CFT integrable system \cite{Beisert:2010jr}.

The {\it third feature} is a consequence of the second feature: when constructing the Q-operators by taking a trace over the oscillator degrees of freedom of the monodromies built from the new types of Lax operators, one needs to include appropriate phase factors $\exp i \Phi_A$ in order to ensure convergence of all matrix elements, as was already stressed in the original BLZ construction \cite{Bazhanov:1998dq}. The physical interpretation of the ``twist angles'' $\Phi_A$ is in terms of Aharonov-Bohm phase factors for the $n$ particles of the system. This slightly breaks the $\alg{su}(n)$ invariance of the chain while being fully compatible with integrability. In fact, the angles should be considered as a beneficial device allowing to expose the intricate integrable ``inner mechanics'' of the model. They may always be considered as a small regulator, which is easily removed from physical observables such as the spectrum. The proper Baxter Q-operators, however, simply do not exist, for good reasons, without these regulating fluxes. It is interesting to observe that similar phase angles also naturally appear in the AdS/CFT context, see \cite{Zoubos:2010kh}, \cite{Arutyunov:2010gu}, \cite{deLeeuw:2010ed}, \cite{Beccaria:2010kd}, \cite{Gromov:2010dy} for the most recent references on this subject.

The final, {\it forth feature} is again closely related to the operatorial construction of the hierarchy of Q-operators constructed from the just discussed previous three features. It is possible to derive certain Q-Q functional relations for the Q-operators. We suspect that the simplest and deepest explanation/derivation for these still remains to be found. Going over to the eigenvalues of the Q-operators, utilizing their analytic structure, which in the case of spin chains is polynomiality in the spectral parameter $z$, and taking certain ratios at critical points in the $z$-plane, one finally derives the spectrum of the chain in the form of the well-known Bethe equations. No ansatz for the wavefunction is made. It is important to stress that the analytic structure of the eigenvalues of the Q-operators, without which the spectrum could not be derived, immediately follows from the trace-over-oscillator-states construction employed. It is {\it not} ad-hoc imposed, as it is (essentially) the case in the current state-of the-art approaches to the AdS/CFT spectral problem, see \cite{Gromov:2010kf} (and in particular the conclusions of this review article), and references therein. 

In the current work we generalize the $\alg{su}(n)$ Q-operator construction of our earlier paper \cite{Bazhanov:2010jq} to the $\alg{su}(n|m)$ case. Our main motivation is again the fact that such spin chains appear in the form of ``closed sectors'' at the one-loop level in the planar AdS/CFT integrable system. In particular, the ``maximally compact closed sector'' has symmetry $\alg{su}(3|2)$ \cite{Beisert:2003ys}, \cite{Beisert:2004ry}, see also \cite{Minahan:2010js}. A smaller closed sector, $\alg{su}(2|1)$, see \cite{Beisert:2005fw}, corresponds at one loop to the famous integrable t-J model, and was first solved by coordinate Bethe ansatz in \cite{Schlottmann:1987zz}, and by algebraic Bethe ansatz in \cite{Essler:1992he, Foerster:1992uk}. The nested Bethe equations of the general $\alg{su}(n|m)$ nearest neighbor magnet were first derived in \cite{Lai:74, Sutherland:1975}, adapting the nested Bethe ansatz method invented, along with the Yang-Baxter equation as well as the scattering interpretation of the Bethe ansatz, for the treatment of the multi-species Bose gas in Yang's seminal paper \cite{Yang:1967bm}. See also \cite{Sutherland:2004aa}. Finally, a further impetus for being interested in supersymmetric Baxter operators, fully consistent with our just spelled out main motivation, is that these have been argued in \cite{Gromov:2010km} to be relevant for the exact solution of the AdS/CFT spectral problem.

We shall find that the construction proceeds, modulo a number of tedious further details mostly related to a multitude of minus signs, very much along the lines of the purely bosonic case. In particular, the above four salient features remain. Now (first feature) one needs to study degenerate solutions of the {\it graded} Yang-Baxter equation (section \ref{sec:YBE}) in order to find the Lax operators needed to build the supersymmetric Baxter operators, and to derive the factorization formulas appropriate to this case (section \ref{sec:fus/fac}). Not surprisingly (second feature), the new Lax operators now contain a mixture of bosonic and fermionic oscillators, cf.~section \ref{sec:YBE}. Similar operators have previously appeared in the literature on the graded, q-deformed systems with $\alg{sl}_q(2|1)$ symmetry \cite{Bazhanov:2008yc}, but here we present the (non-deformed) general $\alg{sl}(n|m)$ derivation from first principles (the Yang-Baxter equation). The (super)trace construction (third feature) of the Q-operators (section \ref{sec:Qconstruction}) proceeds in similarity to the bosonic case, with some amusing differences. We shall find that, for a slightly different reason, the twist angles corresponding to the fermionic particles still need to be non-zero in order to avoid singular equations. In section \ref{sec:oneslashone} we discuss in some detail the complexification $\alg{gl}(1|1)$ of the simplest supersymmetric case $\alg{su}(1|1)$, both for pedagogical reasons as well as a building block of the general case. As concerns the fourth feature, the derivation of the Q-Q equations turns out to be much trickier in the supersymmetric case; we present it in sections \ref{sec:fusion} and \ref{sec:functional}. The Bethe equations (\ref{sec:bethe}) then follow in much the same way as in the bosonic case. We end by illustrating our approach in \ref{sec:tJ} in the well-known case of the t-J model with $\alg{su}(2|1)$ symmetry with its 6 systems of Bethe ansatz equations.

As in the $\alg{su}(n)$ case our method again directly reproduces and {\it explains} the full set of functional equations of the $\alg{su}(n|m)$ spin chain models, which were previously known only on the level of eigenvalues as opposed to operators, see \cite{Kazakov:2007fy}, \cite{Zabrodin:2007rq}, \cite{Tsuboi:2009ud}. Here we would also like to point out that very recently another, apparently both technically and conceptually rather different construction of the $\alg{su}(n|m)$ Q-operators was proposed in the work \cite{Kazakov:2010iu}, albeit with a similar overall motivation. It would be interesting to understand the relation between this work and our approach. There are also a number of interesting papers which deal with the construction of Baxter Q-operators for non-compact spin chains. Apart from the articles cited already in \cite{Bazhanov:2010ts}, \cite{Bazhanov:2010jq}, two notable studies on the non-compact $\alg{sl}(2|1)$ case are \cite{Derkachov:2005hx} and \cite{Belitsky:2006cp}. Again, the precise relation to their approach (if any, it is clear that the method used in these works does not directly apply to compact spin chains) remains to be understood. 

One aspect of our construction method we find particularly appealing is that it {\it naturally} leads to the full underlying combinatorial structure of the nested Bethe ansatz of $\alg{su}(n)$ or $\alg{su}(n|m)$ integrable systems. It was discovered by Z.~Tsuboi \cite{Tsuboi:2009ud} that this structure is best depicted by so-called {\it Hasse diagrams}. These are partially ordered sets of $n$ (or $n|m$) distinguishable objects. In the case at hand, the ordering is given by inclusion. This leads to Hasse diagrams with the connectivity of an $n+m$ dimensional hypercube. The $2^{n+m}$ vertices of the hypercube correspond to all possible subsets of the original $n|m$ objects, and at each such vertex we have exactly one Baxter Q-operator. Thus there are $2^{n+m}$ distinct Q-operators. And indeed, this is also precisely the possible number of degenerate solutions of the (graded) Yang-Baxter equation we find in section \ref{sec:YBE}! There exist $(n+m)!$ different paths starting from the complete set (containing all objects) down to the empty set (containing no objects). This corresponds to the total number of possible nested Bethe ansatz systems. These are all fully equivalent, i.e.~each system leads by itself to the exact spectrum, as long as the twist angles $\Phi_A$ are non-zero. Incidentally, the above mentioned Q-Q relations relevant to the derivation of the Bethe equations also have a very beautiful interpretation: They correspond to the plaquettes (four-cycles) of the hypercube.

In conclusion, the methodology for constructing the complete tower of Baxter Q-operators of compact integrable spin chains with $\alg{su}(n)$ symmetry extends with small but interesting changes to the supersymmetric $\alg{su}(n|m)$ case. The next step will be to include non-compact representations of the quantum space into our framework, and to construct the Baxter Q-operators of the ${\cal N}=4$ one-loop spin chain with $\alg{psu}(2,2|4)$ symmetry, cf.~\cite{Beisert:2003yb}.

 

\section{Graded Permutations and the Graded Yang-Baxter Equation}
\label{sec:YBE}

In this initial section we closely follow section 2 of our previous work \cite{Bazhanov:2010jq}, for the convenience of the reader, and in order to extend our conventions to the graded (and thus supersymmetric) case. The discussion will be terse, and the reader is asked to consult \cite{Bazhanov:2010jq} for ampler explanations. For a nearest neighbor, homogeneous, graded spin chain of length $L$, where each ``spin'' (a misnomer, of course) takes any of the $n+m$ values of the fundamental representation of $\alg{su}(n|m)$, the Hamiltonian reads
\begin{equation}
\label{slnm-ham}
\mathbf{H}_{n|m}
= 
2\sum_{l=1}^L\left(1- \sum_{A,B=1}^{n+m} (-1)^{p(B)}\,e_{AB}^{(l)}
\,e_{BA}^{(l+1)} \right),
\end{equation}
with the superindices taking values $A,B \in 1,2, \ldots, n+m$. Here $e_{AB}$ denotes the $(n+m) \times (n+m)$ matrix unit $(e_{AB})_{ij}=\delta_{Ai}\delta_{Bj}$ and the superscript ``$(l)$'' refers to the quantum space of the $l$-th ``spin'' (better: species of lattice particle) in the chain. The parity function $p(A)$ is a map
\begin{equation}
 p:\{1,\ldots,n+m\}\rightarrow \{0,1\}
\end{equation} 
defining the grading of the vector space\footnote{In the following we will write in slight abuse of notation $(-1)^A$ instead of $(-1)^{p(A)}$.}.
For $A \in \{1,2, \ldots, n\}$ we say that the superindex $A$ takes  ``bosonic'' values, and the grading is defined to be $p(A)=0$. For $A \in \{n+1 \ldots, n+m\}$ the index $A$ takes ``fermionic'' values, and the grading is defined to be $p(A)=1$. As already mentioned in section  \ref{sec:intro}, we will consider flux angles $\Phi_A$, just as in our previous work \cite{Bazhanov:2010jq}. This translates into the following ``quasiperiodic'' boundary conditions:
\begin{equation}
\label{bcH}
e_{AB}^{(L+1)}:= 
e^{i\,(\Phi_A-\Phi_B)}\,e_{AB}^{(1)}
\, ,
\end{equation}
where $\Phi_1,\Phi_2,\ldots,\Phi_{n+m}$ is the set of fixed twist angles. It is easy to show that the Hamiltonian \eqref{slnm-ham} may be rewritten as\footnote{
Actually, in the presence of the fluxes $\Phi_A$ the ``backward'' permutation ${\bf P}_{L,L+1}$ is special, and should include the phase factors of \eqref{bcH}.
}
\begin{equation}
\label{slnm-ham-alt}
\mathbf{H}_{n|m}
= 
2\sum_{l=1}^L\left(1- {\bf P}_{l,l+1} \right),
\end{equation}
where ${\bf P}_{l,l+1}$ denotes the {\it graded} permutation operator on adjacent chain sites $l$, $l+1$. It acts by permuting adjacent species of particles on the lattice, picking up a minus sign iff two fermions are exchanged, i.e.~iff both particles carry a superindex in the set $\{n+1 \ldots, n+m\}$. We see that this Hamiltonian behaves differently when acting on a homogeneous vacuum state. We have
\begin{equation}
\label{bose-fermi-vac}
\mathbf{H}_{n|m} \cdot |B\rangle =0\, ,
\qquad
\mathbf{H}_{n|m} \cdot |F\rangle =4\,L\,|F\rangle\, ,
\end{equation}
where $|B\rangle$ is a ``ferromagnetic'' bosonic vacuum state where any one type of bosonic particle is placed on all lattice sites (there are $n$ such vacua), while $|F\rangle$ is a fermionic vacuum state where any one type of fermionic particle is placed on all lattice sites (there are $m$ such vacua).

Let us now proceed as in  \cite{Bazhanov:2010jq} and derive the general form of linear solutions to the graded Yang-Baxter equation with $\mathfrak{gl}(n|m)$ symmetry\footnote{
From now on we will for simplicity consistently work with the  complexified algebras $\alg{gl}(n|m)$ and $\alg{sl}(n|m)$ instead of the real form $\alg{su}(n|m)$. The quantum space is however in this work always an $L$-fold tensor product of fundamental representations of $\alg{su}(n|m)$.
}. 
To this end we represent the quantum space of the Yang-Baxter equation by the $\mathbb{Z}_2$-graded vector space $\mathbb{C}^{n|m}$ which provides us with the defining relations for the supersymmetric cousin of the previously discussed solutions. Consider the graded Yang-Baxter equation 
\begin{equation}
\label{YB-main}
\Rbf(z_1-z_2)\, \/{\bf L}(z_1) \,
 {\bf L}(z_2)= 
 {\bf L}(z_2)\,\/{\bf L}(z_1)\, \Rbf(z_1-z_2)\,,
\end{equation}
which acts in the space $V\otimes \mathbb{C}^{n|m}\otimes \mathbb{C}^{n|m}$, where $V$ denotes a not yet specified space. Then the intertwiner (R-matrix) $\Rbf(z)$ acts linearly on ${\mathbb C}^{n|m} \otimes {\mathbb C}^{n|m}$ and is defined by
 \begin{equation}\label{Ryang}
{\Rbf}(z)
=z + {\bf P}\, \quad \quad\quad\text{with}\quad\quad\quad {\bf P} =\sum_{A,B}(-1)^B\,e_{AB}\otimes e_{BA}\, ,
\end{equation}
where $ {\bf P}$ is again the just introduced graded permutation, exchanging the states in the two copies of ${\mathbb C}^{n|m} \otimes {\mathbb C}^{n|m}$. It is well known that \eqref{YB-main} serves as the defining relation of the Yangian algebra $Y(\mathfrak{gl}(n|m))$ \cite{molevbook}. More specifically, without loss of generality, choosing the $\bf L$-operators to be of the form
\begin{equation}
{\bf L}(z)=\sum_{A,B}(-1)^{AB+B}L_{AB}(z)\otimes e_{AB}\,,
\end{equation}
the Yang-Baxter equation (\ref{YB-main}) immediately leads to the constraining relations
\begin{equation}\label{Yangiandefinition}
(z_1-z_2)\left[L_{AB}(z_1),L_{CD}(z_2)\right]=(-1)^{AB+AC+BC}\big(L_{CB}(z_2)L_{AD}(z_1)-L_{CB}(z_1)L_{AD}(z_2)\big)\,.
\end{equation} 
Here the brackets denote the supercommutator\footnote{
The supercommutator is given by 
$\left[X,Y\right]=XY-(-1)^{p(X)p(Y)}YX$. The anticommutator will be denoted by $[\cdot,\cdot ]_+$ and the commutator by $[\cdot,\cdot ]_-$.}. 
If we expand $L_{AB}(z)$ in a Laurent series 
\begin{equation}
L_{AB}(z)=L_{AB}^{(0)}+L_{AB}^{(1)}\,z^{-1}+L_{AB}^{(2)}\,z^{-2}+\ldots\,,
\end{equation}
we can rewrite \eqref{Yangiandefinition} in the form
\begin{equation}\label{Yangianpower}
\left[L^{(r)}_{AB},L^{(s)}_{CD}\right]=(-1)^{AB+AC+BC}\sum_{q=1}^{min(r,s)}\bigg(L^{(r+s-q)}_{CB}L^{(q-1)}_{AD}-L^{(q-1)}_{CB}L^{(r+s-q)}_{AD}\bigg)\,.
\end{equation} 
In our discussion we will consider solutions to the graded Yang Baxter equation which are of the form
\begin{equation} \label{Linearsol}                                           
L_{AB}(z)=L^{(0)}_{AB}+z^{-1}L^{(1)}_{AB}\, ,
\end{equation} 
and set all higher terms $L^{(r)}_{AB}=0$ for $r\geq 2$. From (\ref{Yangianpower}) we find that the elements $L^{(0)}_{AB}$ supercommute among themselves, as well as with the elements $L^{(1)}_{AB}$. Therefore we will assume that they are Gra\ss mann numbers. Furthermore, using the GL$(n|m)$ invariance of the R-matrix the entries $L^{(0)}_{AB}$ can be transformed to diagonal form
\begin{equation}
 L^{(0)}_{AB}=\delta_{AI}\,\delta_{BI}\, .
\end{equation} 
 Here $I\subseteq \{1,\ldots,n+m\}$ denotes an arbitrary set containing $|I|$ elements. The only nontrivial commutation relations which arise from (\ref{Yangianpower}) are among the elements $L^{(1)}_{AB}$:
\begin{equation}\label{L1commutation}
  \left[L^{(1)}_{AB},L^{(1)}_{CD}\right] 
=(-1)^{AB+AC+BC}\left(L^{(1)}_{CB}\,L^{(0)}_{AD}-L^{(0)}_{CB}\,L^{(1)}_{AD}\right).
\end{equation} 
In the following 
we assign undotted and dotted indices in order to indicate that they take, respectively, values in the set $I$ and its complement $\bar{I}$. Furthermore, we firstly introduce the $\mathfrak{gl}(I)\equiv\mathfrak{gl}(p|q)$ generators\footnote{Where $p$ and $q$ are the number of elements in $I$ with even and odd grading, respectively. } $E_{AB}$ obeying the usual commutation relations
\begin{equation}\label{slpqgens}
 [E_{AB}\,,\,E_{CD}]=E_{AD}\,\delta_{CB}-(-1)^{(A+B)(C+D)}E_{CB}\,\delta_{AD}\, ,
\end{equation} 
and secondly $|I|\cdot|\bar I|$ pairs of superoscillators, which supercommute with the generators $E_{AB}$, and satisfy
\begin{equation}\label{suposz}
\left[\suposc_{\raisemath{-2pt}{\dot AB}}\,,\,\suposc_{C\dot D}^{\dagger}\right]=\delta_{\raisemath{-2pt}{BC}}\,\delta_{\raisemath{-2pt}{\dot A \dot D}}\,.
\end{equation}
The commutation relations (\ref{L1commutation}) can then be realized with use of the superoscillators (\ref{suposz}) and the $\mathfrak{gl}(p|q)$ generators (\ref{slpqgens}) in the following way:
\begin{eqnarray}
\label{L1solution}
L^{(1)}_{\raisemath{-2pt}{AB}}&=&-(-1)^B\,\left(E_{\raisemath{-4pt}{AB}}+H_{\raisemath{-3pt}{AB}}^I\right)\,;\quad\quad\,
L^{(1)}_{A\dot{B}}=\suposc^{\dagger}_{A\dot{B}}\,;\\
L^{(1)}_{\dot{A}B}&=&-(-1)^B\,\suposc_{\dot{A}B}\,;\qquad\qquad\qquad\quad\,\,
L^{(1)}_{\dot{A}\dot{B}}=\delta_{\dot{A}\dot{B}}\, ,
\end{eqnarray} 
with
\begin{equation}
\label{Hdefinition}
 H_{\raisemath{-3pt}{AB}}^I=\sum_{\dot D\in \bar I}\left(\,\suposc^{\dagger}_{A\dot{D}}\,\suposc_{\raisemath{-4pt}{\dot{D}B}}+\half(-1)^{A+\dot D}\delta_{AB}\right)\,.
\end{equation} 
The definitions above serve as an {\em evaluation homomorphism} of the infinite-dimensional Yangian algebra \eqref{Yangianpower} into a
finite-dimensional algebra composed out of $\alg{gl}(\p|q)$ and the superoscillator algebra defined in (\ref{suposz}). It follows that any representation of this finite-dimensional algebra defines a representation of the Yangian
as well as a solution of the graded Yang-Baxter
equation \eqref{YB-main}. 

For later purposes we arrange the elements $L_{AB}(z)$ in a $2\times 2$ block matrix and define\footnote{The dotted line is \emph{not} separating fermionic entries from bosonic ones!}
 \begin{equation}\label{Lcanon}
\Lbf_I(z)=\left(\begin{BMAT}(@,30pt,30pt)
{c.c}{c.c}
z\,\delta_{\raisemath{-4pt}{AB}}-(-1)^B\,\left(E_{\raisemath{-4pt}{AB}}+H_{\raisemath{-3pt}{AB}}^I\right)&\suposc^{\dagger}_{A\dot{B}}\\
-(-1)^B\,\suposc_{\dot{A}B}&\delta_{\dot A\dot B}
\end{BMAT}
\right)\,.
\end{equation}
We will refer to (\ref{Lcanon}) as the \textit{linear canonical} $\Lbf$-operator. Any first order $\Lbf$-operator with $L^{(0)}_{AB}$ of rank $|I|$ and with non-degenerate $L^{(1)}_{\dot A\dot B}$ can be brought to this form using the aforementioned GL$(n|m)$ invariance.
 

\section{Fusion and Factorization of L-operators}
\label{sec:fus/fac}

An essential part of our analysis in the following is based 
on some remarkable decomposition properties of the product of two
$\Lbf$-operators of the form \eqref{Lcanon}. 
The Yangian ${\mathcal Y}= Y(\alg{gl}(n|m))$
is a Hopf algebra, see e.g.~\cite{molevbook}. 
In particular, its co-multiplication 
\beq
{\mathcal Y}\to {\mathcal Y}\otimes 
{\mathcal Y}
\label{comul}
\eeq
is generated by the matrix product 
of two $\Lbf$-operators, corresponding to two different copies of
${\mathcal Y}$ appearing on the RHS of \eqref{comul}.
We are interested in the structure of the product 
\beq
\Lbf(z)=\Lbf_I^{[1]}(z+\omega_1)\,\Lbf_J^{[2]}(z+\omega_2)\,,
\label{coprod}
\eeq
where suffices $[1]$ and $[2]$ have been added to emphasize that the matrix entries of the corresponding $\Lbf$-operators act on different spaces, and in consequence supercommute. 

The quantity \eqref{coprod} for two non-intersecting sets $I\cap J=\varnothing$ will be considered
in section \ref{sec:factorization}.
In this case the product \eqref{coprod} is linear in the spectral parameter $z$ and belongs to the family of solutions \eqref{Linearsol} studied in the previous section. Following the same reasoning as in \cite{Bazhanov:2010jq}, this will lead to a new instance of the remarkable \emph{factorization} properties of the $\Lbf$-operators we found in our earlier work.
In turn, in section \ref{sec:fusion} the product \eqref{coprod} is considered for the case in which $I\cap J\neq\varnothing$. This case had not been discussed earlier in \cite{Bazhanov:2010jq}. Our analysis will lead to a simple and elementary derivation of an important set of functional relations  in section \ref{sec:functional}.

\subsection{Fusion: Non-Intersecting Sets}
\label{sec:factorization}
The procedure described below is a generalization of the one presented in \cite{Bazhanov:2010jq}. 
Let us start by taking $I$ and $J$ to be two non-intersecting sets. By permuting rows and columns one can rewrite the  $\Lbf_I(z)$ and $\Lbf_J(z)$ operators in the following way 
 \begin{equation}
\Lbf^{[1]}_I(z)=\left(\begin{BMAT}(@,30pt,30pt,30pt)
{c.c.c}{c.c.c}
z\,\delta_{\raisemath{-4pt}{AB}}-(-1)^B\,\left(E^{[1]}_{\raisemath{-4pt}{AB}}+H_{AB}^{[1]\,I}\right)&\suposc^{{\dagger}\,[1]}_{A\dot{B}}&\suposc^{{\dagger}\,[1]}_{A\ddot{B}}\\
-(-1)^B\suposc^{[1]}_{\dot{A}B}&\delta_{\dot A\dot B}&0\\
-(-1)^B\suposc^{[1]}_{\ddot{A}B}&0&\delta_{\ddot A\ddot B}
\end{BMAT}
\right)
\end{equation}
and
 \begin{equation}
\Lbf^{[2]}_J(z)=\left(\begin{BMAT}(@,30pt,30pt,30pt)
{c.c.c}{c.c.c}
\delta_{AB}&-(-1)^{\dot B}\,\suposc^{[2]}_{A\dot B}&0\\
\suposc^{{\dagger}\,[2]}_{\dot A{B}}&z\,\delta_{\raisemath{-4pt}{\dot A\dot B}}-(-1)^{\dot B}\,\left(E^{[2]}_{\raisemath{-4pt}{\dot A\dot B}}+H_{\dot A\dot B}^{[2]\,J}\right)&\suposc^{{\dagger}\,[2]}_{\dot A\ddot{B}}\\
0&-(-1)^{\dot B}\,\suposc^{[2]}_{\ddot{A}\dot B}&\delta_{\ddot A\ddot B}
\end{BMAT}
\right)\,,
\end{equation}
where 
\begin{equation}
 A,B,C \in I\,,\qquad \dot{A},\dot{B}, \dot{C} \in J\,,\qquad \ddot{A},\ddot{B}, \ddot{C} \in \overline{I \cup J}\,.
\end{equation}
As extensively discussed in \cite{Bazhanov:2010jq}, the co-product of $\Lbf^{[1]}_I(z)$ and $\Lbf^{[2]}_J(z)$ generates a solution $\Lbf_{I\cup J}(z)$ to the Yang-Baxter equation. This is also valid for the graded Yang-Baxter equation. One finds
\begin{equation}\label{fusionLL}
 \Lbf^{[1]}_I\bigg(z+\half\sum_{\dot D\in J}(-1)^{\dot D}\bigg)\,\Lbf^{[2]}_J\bigg(z-\lambda-\half\sum_{D\in I}(-1)^D\bigg)=\mathcal{S}\,\Lbf_{I\cup J}(z)\,G\,\mathcal{S}^{-1}\,,
\end{equation}  
which is a rather remarkable factorization formula. The similarity transform 
\begin{equation}
 \mathcal{S}=\exp \left[\sum_{\raisemath{-2pt}{A\in I}}\,\sum_{\dot B\in J}\,\sum_{\ddot C\in\overline{I\cup J}}\suposc^{{\dagger}\,[1]}_{ A{\dot B}}\left((-1)^A\,\suposc^{{\dagger}\,[2]}_{{\dot B}A}+\suposc^{{\dagger}\,[2]}_{{\dot B}\ddot C }\,\suposc^{[1]\phantom{\dagger}}_{\raisemath{-4pt}{{\ddot C}A}}\right)\right]\,,
\end{equation}  
and the $z$ independent matrix\footnote{Note that $\suposc_{A\dot B}$ contained in $G$ supercommute with the elements of $\Lbf_{I\cup J}(z)$.}
\begin{equation}
 G=\left(\begin{BMAT}(@,20pt,25pt,20pt)
{c.c.c}{c.c.c}
\delta_{AB}&-(-1)^{\dot B}\,\suposc^{[2]}_{A\dot B}&0\\
0&\delta_{\dot A\dot B}&0\\
0&0&\delta_{\ddot A\ddot B}
\end{BMAT}
\right)\,
\end{equation}
have been introduced to  write $\Lbf_{I\cup J}(z)$ in the canonical form (\ref{Lcanon}) 
\begin{equation}
\Lbf_{I\cup J}(z)=\left(\begin{BMAT}(@,50pt,40pt)
{c.c}{c.c}
z\,\delta_{\hat A\hat B}-(-1)^{\hat B}\left(\tilde E_{\hat A\hat B}+H_{\hat A\hat B}^{I\cup J}\right)&\suposc^{\dagger}_{\hat A\ddot{B}}\\
-(-1)^{\hat B}\,\suposc_{\ddot{A}\hat B}&\delta_{\ddot A\ddot B}
\end{BMAT}
\right).
\end{equation}
Hatted indices take values from the merged ordered set $I\cup J$, i.e. $\hat A=(A,\dot A)$. The objects $\tilde E_{\hat A\hat B}$ obey  $\mathfrak{gl}(I\cup J)$ commutation relations and are of the form
\begin{equation}
\label{inducedrep}
 \begin{split}
\tilde{E}_{AB}&=E^{[1]}_{AB}+\suposc^{\dagger\,[1]}_{A\dot{C}}\suposc^{[1]}_{\raisemath{-4pt}{\dot{C} B}}\\
\tilde{E}_{A\dot{B}}&=(-1)^{\dot B}\suposc^{\dagger\,[1]}_{A\dot{B}}\,\lambda-(-1)^{(\dot B+\dot D)(\dot B+C)}\suposc^{\dagger\,[1]}_{ A{\dot D}}\suposc^{\dagger\,[1]}_{C{\dot B}}\suposc^{[1]}_{\raisemath{-4pt}{{\dot D}C}}+\suposc^{\dagger\,{[1]}}_{A\dot{C}}\,E^{[2]}_{\dot C\dot B}-(-1)^{\dot B
+ C}E^{[1]}_{AC}\,\suposc^{\dagger\,[1]}_{C\dot{B}}\\
\tilde{E}_{\dot{A}B}&=\suposc^{[1]}_{\dot{A}B}\\
\tilde{E}_{\dot A\dot B}&=E^{[2]}_{\dot A\dot B}+\lambda\,(-1)^{\dot B}\delta_{\dot A\dot B}-(-1)^{(\dot A+ \dot B)(\dot B+ C)}\,\suposc^{\dagger\,{[1]}}_{C\dot{B}}\,\suposc^{[1]}_{\raisemath{-4pt}{\dot{A} C}}\,,
 \end{split}
\end{equation} 
where summation is understood to be over the range of the repeated indices.
\subsection{Fusion : Intersecting Sets}
\label{sec:fusion}

In this section we consider products of the form \eqref{coprod} for general non-intersecting sets $I$, $J$ and $K$. Namely,
\begin{equation}
\label{prodlax}
 \Lbf^{[1]}_{I\cup J}(z+\omega_1)\,\Lbf^{[2]}_{I\cup K}(z+\omega_2)\,.
\end{equation}
In particular we are interested in the relation between 
\begin{equation}
 \Lbf^{[1]}_{I\cup J}(z+\omega_1)\,\Lbf^{[2]}_{I\cup K}(z+\omega_2)\, \qquad \text{and}\qquad \Lbf^{[1]}_{I\cup J'}(z+\omega'_1)\,\Lbf^{[2]}_{I\cup K'}(z+\omega'_2)\,,
\end{equation}
for $J\cup K= J'\cup K'$. 
This analysis leads to a derivation of an important set of functional relations known as ${\bf Q}$-$ {\bf Q}$ relations\footnote{To avoid misunderstandings, we recall that in the literature another set of functional relations is sometimes referred to as ${\bf Q}$-$ {\bf Q}$ relations. In this paper {\bf Q} will always refer to Baxter's $\Qop$-operators.}.
For a discussion on this point and more functional relations see section \ref{sec:functional}.

Let us us take a closer look at \eqref{prodlax}. If the set $I$ is not empty, this product takes the form
\begin{equation}
z^2\,\tilde{L}^{(0)}+z\,\tilde{L}^{(1)}+\,\tilde{L}^{(2)}\,,
\end{equation}
and as such does not fit into the classification of Lax operators as written in \eqref{Linearsol}. To analyze this more complicated Lax operator, it is convenient to directly restrict the analysis to only a part of the structure of \eqref{prodlax}. It will be argued that the remaining structure is then uniquely fixed by the fact that \eqref{prodlax} is a solution to the Yang-Baxter equation.
For simplicity and for the purposes of section \ref{sec:functional} we will consider $\Lbf_{I\cup J}$ and $\Lbf_{I \cup K}$  with $E\equiv 0$ (see equation \eqref{Lcanon}), the general case can then be analyzed in a similar way.
The product \eqref{prodlax} can be conveniently rewritten as
\begin{equation}
\label{quadspec}
 \mathcal{S}\,\left(z^2\,L^{(0)}+z\,L^{(1)}+\,L^{(2)}\right)\,G\,\mathcal{S}^{-1}\,,
\end{equation}
with
\begin{equation}
\label{L0fusion}
L^{(0)}=\left(
\begin{BMAT}(r)[0.15cm,0cm,0cm]{c.c.c}{c.c.c}
\delta_{AB} &\ 0 &\ 0\\
0&0&0\\
0&  0& 0 
\end{BMAT}\, \right)\ 
\end{equation}
\begin{equation}
\label{L1fusion}
L^{(1)}=\left(
\begin{BMAT}(r)[0.15cm,0cm,0cm]{c.c.c}{c.c.c}
-\left(-1 \right)^{B}\,J_{AB} &\ \suposc_{A\dot B}^\dagger\ &\ \suposc_{A\ddot B}^\dagger \\
-(-1)^{ B}\,\suposc_{\dot{A} B}&\delta_{\dot A \dot B}&0\\
-(-1)^{ B}\,\tilde\suposc_{\ddot{A} B}\, \, &  0 \, \,& \, \, 0
\end{BMAT}\, \right)\ 
\end{equation}
\begin{equation}
\label{L1fusionbis}
J_{AB}=H^I_{AB}+\tilde H^{{ {I\cup J}}}_{AB}-\left(\omega_1+\omega_2\right)\,\delta_{AB}\,\left(-1\right)^B\,,
\end{equation} 
\begin{equation}
\label{L2fusion}
L^{(2)}_{\ddot A \ddot B}= \delta_{\ddot A \ddot B}-(-1)^{C}\,\tilde\suposc_{\ddot{A} C}\,\suposc_{C\ddot B}^\dagger\,,
\end{equation}
%
%
%
%
%
where $A,B,C \in I$, $\dot{A}, \dot{B},\dot{C} \in J \cup K $ and $\ddot{A}, \ddot{B}, \ddot{C} \notin  I \cup J \cup K$. The operator $H$ is defined in \eqref{Hdefinition}.
The similarity transform $\mathcal{S}$ and the matrix $G$ are given in  appendix \ref{app:details} together with the identification of the oscillators from \eqref{prodlax}, and the ones appearing in \eqref{L1fusion}. The analysis of 
\eqref{prodlax} is greatly simplified by the following observation:

\hspace{0.5cm}

\noindent
\textbf{Proposition:} 
{\it The Yang-Baxter equation \eqref{YB-main} for 
\begin{equation}\label{Lsquared}
\Lbf_{I^2\cup J \cup K}(z)\equiv\left(z^2\,L^{(0)}+z\,L^{(1)}+\,L^{(2)}\right),
\end{equation}
together with
 \eqref{L0fusion}, \eqref{L1fusion}, \eqref{L1fusionbis},  \eqref{L2fusion} fixes all the entries of $L^{(2)}$ uniquely\footnote{The uniqueness is up to algebra automorphisms. In the present construction they manifest themselves as similarity transforms $\mathcal S$.}
 up to the choice of $\alg{gl}(J \cup K)$ generators $E_{\dot{A} \dot{B}}$. For this reason it will be denoted by
\begin{equation}
\Lbf^{\mathfrak{Rep}}_{I^2\cup J \cup K}(z|\,\omega_1+\omega_2)\,,
\end{equation}
where $\mathfrak{Rep}$ denotes some representation of the $\alg{gl}(J\cup K)$ algebra.
 The entries of $\Lbf^{\mathfrak{Rep}}_{I^2\cup J \cup K}$ belong to the direct product of the universal enveloping algebra of families of superoscillator algebra 
and  $\alg{gl}(J \cup K)$ generators $E_{\dot{A} \dot{B}}$.  }

\hspace{0.5cm}

\noindent
A detailed proof of the statement above and the analysis of related structures will be presented in a separate work. Let us stress a simple but important part of the derivation. 
 On general grounds the Yangian algebra contains Yangian subalgebras. In the present paper this property takes the form
\begin{equation}
 Y(\alg{gl}(n|m)) \supset Y(\alg{gl}(I))\otimes Y(\alg{gl}(\bar{I})).
\end{equation}
A closer look at \eqref{quadspec}  and \eqref{L0fusion} immediately reveals that the representation of the Yangian subalgebra $Y(\alg{gl}(\bar{I}))$ is of the type \eqref{Linearsol}, being a linear function of the spectral parameter. It fits in the classification scheme of section \ref{sec:YBE}. For this reason one concludes that\footnote{Compare with \eqref{L1solution} and \eqref{Hdefinition}. }
\begin{equation} 
\label{L2easy}
 L^{(2)}_{\dot A\ddot B}= \suposc_{\dot A\ddot B}^\dagger\,,\qquad  L^{(2)}_{\ddot A\dot B}= -(-1)^{\dot B}\,L^{(2)}_{\ddot A \ddot C}\,\,\suposc_{\ddot{C}\dot B}\,, 
\end{equation}
\begin{equation}
\label{L2easybis}
L^{(2)}_{\dot A\dot B}=-\left(-1\right)^{\dot{B}}\left(E_{\dot{A} \dot{B}}+\sum_{\ddot{C}}\left(\suposc^{\dagger}_{\dot A \ddot{C}}\,\suposc_{\ddot{C}\dot B}+\half\left(-1\right)^{\dot{A}+\ddot{C}}\,\delta_{\dot{A}\dot{B}}\right)\right),
\end{equation}
where $E_{\dot{A}\dot{B}}$ are $\alg{gl}(J \cup K)$ generators and, together with the superoscillators   $(\suposc_{\ddot{A}\dot B},\, \suposc^{\dagger}_{\dot A \ddot{B}})$,  supercommute with all the elements of \eqref{L1fusion}. 
The involved part of the derivation of the proposition above consists in showing that all the other entries of $L^{(2)}$
\begin{equation}
\label{L2hard}
 L^{(2)}_{\dot A B}\,, \qquad  L^{(2)}_{\ddot A B}\,,\qquad L^{(2)}_{A \dot B}\,, \qquad  L^{(2)}_{A \ddot B}\,, \qquad L^{(2)}_{A B}\,,
\end{equation}
are uniquely fixed by the Yang-Baxter equation. As stated above, the detailed forms of the quantities in \eqref{L2hard} is built from oscillators and the generators $E_{\dot A \dot B}$. They do not contain new degrees of freedom with respect to \eqref{L1fusion} and \eqref{L2easy}. This part of the derivation will be omitted.


The structure just described comes from the Yang-Baxter equation. Using this insight as a guiding principle 
one can arrange \eqref{prodlax} in the form stated in the proposition above choosing $\mathcal{S}$ and $G$ appropriately (appendix \ref{app:details}) in \eqref{quadspec}. An explicit computation fixes the form of the $\alg{gl}(J \cup K)$ generators $E_{\dot{A}\dot{B}}$. The realization of $\alg{gl}(J \cup K)$ is a special case of \eqref{inducedrep}, for convenience we rewrite it here  \footnote{The oscillators that realize $E_{\dot{A} \dot{B}}$ are not the one explicitly appearing in  \eqref{L1fusion} and \eqref{L2easy}.} 
\begin{equation}
\label{slstuckedgen}
 \begin{split}
E_{\dot{A}_1\dot{B}_1}&=\suposc^{\dagger\,[1]}_{\dot{A_1}\dot{C_2}}\suposc^{[1]}_{\raisemath{-4pt}{\dot{C}_2 \dot{B}_1}}-\alpha_1 \left(-1\right)^{\dot{B}_1}\,\delta_{\dot{A}_1\dot{B}_1}\,\\
 E_{\dot{A}_1\dot{B}_2}&=-\suposc^{\dagger\,[1]}_{\dot{A}_1\dot{C}_2}\left((-1)^{(\dot{B}_2+\dot{C_2})(\dot{B}_2+\dot{C}_1)}\suposc^{\dagger\,[1]}_{ \dot{C}_1\dot{B}_2}\,\suposc^{[1]}_{\raisemath{-4pt}{\dot{C}_2 \dot{C}_1}}-(\alpha_1-\alpha_2)\,\left(-1\right)^{\dot{C}_2}\delta_{\dot{C}_2\dot{B}_2} \right)\\
E_{\dot{A}_2\dot{B}_1}&=\suposc^{[1]}_{\dot{A}_2\dot{B}_1}\\
E_{\dot{A}_2\dot{B}_2}&=-(-1)^{(\dot{A}_1+ \dot{B}_1)(\dot{B}_1+ \dot{C}_2)}\,\suposc^{\dagger\,{[1]}}_{\dot{C}_2\dot{B}_1}\,\suposc^{[1]}_{\raisemath{-4pt}{\dot{A}_1 \dot{C}_2}} - \alpha_2\left(-1\right)^{\dot{B}_2}\,\delta_{\dot{A}_2\dot{B}_2}\,
 \end{split}
\end{equation} 
\begin{equation}
\label{alphadef}
 \alpha_1\equiv\omega_1-\sfrac{1}{2}\sum_{\dot{D}_2 \in K}\left(-1\right)^{\dot{D}_2}\,,\qquad \alpha_2\equiv\omega_2+\sfrac{1}{2}\sum_{\dot{D}_1 \in J}\left(-1\right)^{\dot{D}_1}\,,
\end{equation}
where sums over repeated indices are understood, and $\dot{A}_1, \dot{B}_1, \dot{C}_1 \in J$, $\dot{A}_2, \dot{B}_2, \dot{C}_2 \in K$. To summarize:
For any two non intersecting sets $J$,$K$ such that the set $J\cup K$ and the quantity $\omega_1+\omega_2 $ are fixed, the product \eqref{prodlax} takes the same form up to appropriate $\mathcal S$ and $G$ (see appendix \ref{app:details}) and $\alg{gl}(J \cup K )$ generators given by  
\eqref{slstuckedgen}. 

A particularly interesting case of the fusion considered in this section is the one in which $J \cup K$ contains only two elements:
\begin{equation}
 J \cup K=\{A,B\}\,,
\end{equation}
this can happen in two inequivalent ways, namely
\begin{equation}
 (i):\,\qquad J=\{A\}\,,\qquad K=\{B\}\,,
\end{equation}
\begin{equation}
 (ii):\,\qquad J=\{A,B\}\,, \qquad K=\varnothing\,.
\end{equation}
The result above reads respectively
\begin{equation}
\label{fusion1}
 (i):\qquad \Lbf^{[1]}_{I\cup\, A }(z+\omega_1)\,\Lbf^{[2]}_{I\cup\, B}(z+\omega_2)\sim \Lbf^{\pi^+_{\Lambda}}_{I^2\cup\, A \cup\, B}(z|\,\omega_1+\omega_2)\, G\,,
\end{equation}
\begin{equation}
\label{fusion2}
 (ii):\qquad\Lbf^{[1]}_{I\cup\, A \cup\, B }(z+\omega'_1)\,\Lbf^{[2]}_{I}(z+\omega'_2)\sim \Lbf^{\text{singlet}}_{I^2\cup \,A \cup\, B}(z|\,\omega'_1+\omega'_2)\,,
\end{equation}
the symbol $\sim$ relates quantities that differ by similarity transform acting only in the oscillator space. The explicit form of the $\alg{gl}(\{A,B\})$ generators in \eqref{fusion1} and \eqref{fusion2}, denoted by $\pi^{+}_\Lambda$ and \emph{singlet} respectively, can be obtained specializing
the expressions \eqref{slstuckedgen} and \eqref{alphadef}.
For convenience we write them explicitly in the following:

If $p(A)=p(B)$, $\alg{gl}(\{A,B\})=\alg{gl}(2)$  with generators\footnote{These are $E_{\dot{A}\dot{B}}$ generators written in the same order as in \eqref{slstuckedgen}.}
\begin{equation}
\label{gl2}
 (i):\,\,
\begin{cases}
\osca^\dagger\, \osca +\half -(-1)^{p(A)}\,\omega_1\,,\\
\osca^{\dagger}\left((-1)^{p(A)}\,(\omega_1-\omega_2)-1 -\osca^\dagger\, \osca\right)\,,\\
\osca\,,\\
-\osca^\dagger\, \osca -\half -(-1)^{p(A)}\,\omega_2\,,
\end{cases}\qquad
 (ii):\,\,
\begin{cases}
-(-1)^{p(A)}\,\omega'_1\,,\\
0\,,\\
0\,,\\
-(-1)^{p(A)}\,\omega'_1\,,
\end{cases}
\end{equation}

If $p(A)\neq p(B)$, $\alg{gl}(\{A,B\})=\alg{gl}(1|1)$  with generators 
\begin{equation}
\label{gl11}
 (i):\,\,
\begin{cases}
\oscc^\dagger\, \oscc - \half -(-1)^{p(A)}\,\omega_1\,,\\
(-1)^{p(A)}\,\left(\omega_1-\omega_2\right)\oscc^{\dagger}\,,\\
\oscc\,,\\
-\oscc^\dagger\, \oscc +\half +(-1)^{p(A)}\,\omega_2\,,
\end{cases}\qquad
 (ii):\,\,
\begin{cases}
-(-1)^{p(A)}\,\omega'_1\,,\\
0\,,\\
0\,,\\
(-1)^{p(A)}\,\omega'_1\,,
\end{cases}
\end{equation}
where $(\osca,\osca^\dagger)$ and $(\oscc,\oscc^\dagger)$ are bosonic and fermionic oscillators respectively. The results from this sections will be used in section \ref{sec:functional}.

\section{Construction of the Q-operators}
\label{sec:Qconstruction}

The purpose of this section is to construct $\bf T$- and $\bf Q$-operators.
They form a family of operators commuting with the Hamiltonian \eqref{slnm-ham}.
These operators act on the quantum space
which is an $L$-fold tensor product\footnote{We define a tensor product as $X\otimes Y =(-1)^{(A+B)(C+D)}X_{AB}\,Y_{CD}\,e_{AB}\otimes e_{CD}$.} of the fundamental representations 
of the algebra $\mathfrak{gl}(n|m)$,
\begin{equation}\label{quantumspace}
\underbrace{
{\mathbb C}^{n|m}\otimes {\mathbb C}^{n|m}\otimes \cdots \otimes
{\mathbb C}^{n|m}}_{L-\mbox{\scriptsize{times}}}
\ .
\end{equation}
In this representation solutions of the Yang-Baxter equation 
\eqref{YB-main} are $(n+m) \times (n+m)$ matrices, acting in the
quantum space of a single spin. Their matrix elements are operators   
in some representation space $V$ of the Yangian algebra $Y(\mathfrak{gl}(n|m))$.
This representation space will be called the auxiliary space. 
For each solution of \eqref{YB-main} one can define a transfer matrix
\begin{equation}\label{gen.transfer.matrix}
 {\mathbb T}_{V}(z)=\mbox{Str}_{V} \Big\{{ \Dbf}\, {\Lbf}(z)\otimes
 {\Lbf}(z)\otimes \cdots \otimes{\Lbf}(z)\Big\}.
 \end{equation}
The tensor product in \eqref{gen.transfer.matrix} is taken in the quantum spaces
${\mathbb C}^{n|m}$, while the operator product and the trace is taken
with respect to the auxiliary space $V$. The quantity 
${\mathbb D}$ is a {\it boundary twist} operator acting only in
the auxiliary space, i.e.~it acts trivially in the quantum space.
This boundary operator is completely determined by the requirement of
commutativity of the 
transfer matrix \eqref{gen.transfer.matrix} with the Hamiltonian \eqref{slnm-ham},
which leads to the following conditions
\begin{equation}\label{D-def}
 { \Dbf}\,\big({\Lbf}(z)\big)_{AB} \,{\Dbf}^{-1}=
 e^{i\,\left(\Phi_B-\Phi_A\right)}\, \big( {\Lbf}(z)\big)_{AB}\,, 
 \qquad A,B=1,\ldots,n+m\,.
 \end{equation}
Solving the latter for the $\Lbf$-operator \eqref{Lcanon} with the
arbitrary set 
$I$,
 one obtains 
\begin{eqnarray}\label{D-exp}
\Dbf_{I}&=&\exp\Big\{-i\sum_{A\in I} \Phi_A E_{AA}-
i\sum_{A,B} (\Phi_A-\Phi_{B}) \,\suposc^{\dagger}_{AB}
\suposc_{BA}\Big\}\,,
\end{eqnarray}
where the last summation is over all oscillators present in $\Lbf_{I}$.

 Consider now the most general
$\Lbf$-operator \eqref{Lcanon} with an arbitrary set $I$.
 Recall that the
matrix elements of \eqref{Lcanon} belong to the direct product of the algebra $\alg{gl}(I)$ and  $|I|\cdot |\bar{I}|$ 
copies of superoscillator algebras. 
We therefore have to define the supertrace over both the superoscillator representation space as well as over some $\alg{gl}(I)$ module. 
As stressed in \cite{Bazhanov:2010jq}, the supertrace is completely determined by the commutation
relations \eqref{suposz}, definition \eqref{Hdefinition} and the cyclic property of the supertrace, the specific choice of
the representations is not important as long as the supertrace exists. It is convenient, however, for the purpose of direct calculations, to specify the superoscillator algebra representation. For bosonic oscillators\repfoot $(\osca^\dagger,\osca)$ we take the infinite-dimensional Fock representation spanned by the vectors $|k\rangle$, $k=0,1,\ldots,\infty$ which are defined by 
\begin{equation}
\osca|0\rangle=0\,,\qquad \osca^{\dagger}|k\rangle=|k+1\rangle\,.
\end{equation} 
For fermionic oscillators\repfoot $(\oscc^{\dagger},\oscc)$ we take the two-dimensional representation spanned by the vectors $|\bar 0\rangle,|\bar 1\rangle$ and defined by
\begin{equation}\label{fermionic.rep}
\oscc|\bar 0\rangle=0\,,\quad \oscc^{\dagger}|\bar 1 \rangle=0\,,\quad  \oscc^{\dagger}|\bar 0 \rangle=|\bar 1\rangle\,,\quad  \oscc|\bar 1 \rangle=|\bar 0\rangle\,.
\end{equation}
Let $P(\suposc,\suposc^{\dagger})$ be an arbitrary polynomial of the
superoscillators $\suposc$ and $\suposc^{\dagger}$. 
Below it will be convenient to use a {\em normalized supertrace} over the
representation ${\mathcal F}$,
\beq
\nstr_{\mathcal F} \Big\{e^{i\Phi \osch}
P(\suposc,\suposc^\dagger)\Big\}\ \ \mathop{=}^{\mbox{\small 
def}}\ \ \frac { \mbox{Str}_{\mathcal F} \Big\{e^{i\Phi\osch}
P(\suposc,\suposc^\dagger)\Big\}_{\phantom{|}}}{\mbox{Str}_{\mathcal F} 
\Big\{e^{i\Phi\osch}\Big\}^{\phantom{|}}}\ , 
\label{norm-tr}
\eeq
where $\mbox{Str}_{\mathcal F}$ denotes the standard supertrace.  On the other hand, we will not spell out the $\alg{gl}(I)$ module, the notation   $\mathfrak{Rep}$ will be used to label some unspecified choice. Of course the representation $\mathfrak{Rep}$  has to be chosen such that the supertrace exists. For the purpose of this paper we will only use highest weight representations.


We are now ready to define various transfer matrices, all commuting
with the Hamiltonian \eqref{slnm-ham} and among each other.  Substituting \eqref{Lcanon} and \eqref{D-exp} into
\eqref{gen.transfer.matrix} one can define rather general transfer matrices
\beq
{\Xbf}^{\mathfrak{Rep}}_{I}(z)=
e^{i z\,(\,\sum_{A\in I}(-1)^{A}\Phi_A)
}\ \mbox{Str}^{\alg{gl}(I)}_{  \mathfrak{Rep}}\,\,
\nstr_{{\mathcal F}^\star} 
\big\{\/{\bf M}_{I}(z)\big\}\,,\label{Z-def}
\eeq
where ${\bf M}_{I}(z)$ is the corresponding monodromy matrix, 
\beq
\qquad {\bf M}_{I}(z)=
{\Dbf}_{I}\, {\mathbf L}_{I}(z)\otimes
 {\mathbf L}_{I}(z)\otimes\cdots \otimes{\mathbf L}_{I}(z)\,.\label{M-def}
\eeq
Here $\nstr_{{\mathcal F}^\star}$ denotes the normalized supertrace
\eqref{norm-tr} over all involved oscillator representations, 
while $\mbox{Str}_{\mathfrak{Rep}}$ denotes the supertrace over our chosen but unspecified representation of $\alg{gl}(I)$.  The exponential scalar factor in front of the supertrace is introduced for later convenience\footnote{The overall normalization of transfer matrices is an interesting issue. For example, the universal R-matrix approach leads to a normalization involving spectral parameter dependent ratios of gamma functions such that the R-matrices satisfy certain crossing relations, see e.~g.~\cite{Rej:2010mu} and references therein.}.  
For the constructions of the present paper it is natural to distinguish  some ${\Xbf}^{\mathfrak{Rep}}_{I}(z)$
in the family \eqref{Z-def}. The operator \eqref{Z-def} will be denoted by
\beq
\label{Xplus}
{\Xbf}_{I}^+(z,\Lambda_{I})
\eeq
where $\mathfrak{Rep}$ is now an infinite-dimensional highest
weight representation (Verma-module) $\pi_{\Lambda_I}^{{+}}$.

The monodromy matrices ${\Xbf}_{I}^+(z,\Lambda_{I})$ and ${\Xbf}^{\mathfrak{Rep}}_{I}(z)$ for a given
$\mathfrak{Rep}$ are related\footnote{All the Casimir operators have to take the same values in $\mathfrak{Rep}$ and $\pi_{\Lambda_I}^{{+}}$.}. The relation between the two is exactly the same as the one between the $\alg{gl}(I)$ characters  over the corresponding modules. In the case of finite dimensional representations of the $\alg{gl}(n)$ algebra this relation is nicely encoded in the BGG result \cite{BGG}. This result has been used in \cite{Bazhanov:2010jq} to derive functional relations among transfer matrices.
For the  $\alg{gl}(n|m)$ superalgebra the relation between infinite dimensional Verma modules
and finite dimensional representations has apparently been extensively studied, but the results are less transparent compared to the $\alg{gl}(n)$ case due to atypical representations\footnote{It is worth pointing out that a mechanism analog to atypicality exists also for some infinite dimensional representations of $\alg{gl}(n)$. An example is the conserved current multiplet of the four dimensional conformal algebra.}.
More comments on this point are postponed to section \ref{sec:functional}.


As stressed before the $\Xbf$-operators defined above are rather general transfer matrices.
Two limiting cases of these operators are particularly relevant. If one takes $I$ to be the full set
\eqref{Z-def} reduces to the standard $\Top$-operator
\begin{equation}
\Top_{\mathfrak{Rep}}(z)\equiv \Xbf^{\mathfrak{Rep}}_{\{1,\dots,n+m\}}(z)=\mbox{Tr}_{\mathfrak{Rep}}\left\{ \Dbf \mathcal L(z) \otimes \mathcal L(z)\otimes\ldots\otimes \mathcal L(z)  \right\}
\end{equation}
where $\mathcal L(z)=\Lbf_{\{1,\ldots,n+m\}}(z)$ and $\mathfrak{Rep}$ is some representation of $\alg{gl}(n|m)$.
 The boundary operator reduces to
\begin{equation}
\Dbf=\Dbf_{\{1,2,\ldots,n,n+1,\ldots,n+m\}}=\exp\left(i\sum_{A=1}^{n+m}\Phi_AE_{AA}\right).
\end{equation}
The other limit corresponds to the trivial one-dimensional representation of $\mathfrak{gl}(I)$. The resulting operators are called $\Qop$-operators
\begin{equation}
\Qop_{I}(z)=\Xbf^{\text{singlet}}_{I}(z).
\end{equation}
The $\Qop$-operators are labeled by the set $I$. There are $2^{n+m}$ such sets, and therefore the same number of $\Qop$-operators. As already stressed in section \ref{sec:intro}, the $\Qop$-operators can be conveniently associated with the nodes of a hypercubical Hasse diagram with order given by inclusion on the sets $I$. For more on this, see  section \ref{sec:functional}.


\section{ \texorpdfstring{$\mathfrak{gl}(1|1)$}{}}
\label{sec:oneslashone}
Before proceeding to the derivation of functional relations among the transfer matrices constructed in the previous section for $\mathfrak{gl}(n|m)$ spin chains,  we will analyze in this section the $\mathfrak{gl}(1|1)$ example. This case in conjunction with the $\mathfrak{gl}(2)$ case serve as building blocks for the higher rank  $\mathfrak{gl}(n|m)$ algebras. In the following we will put particular emphasis on the differences between the $\mathfrak{gl}(1|1)$ and $\mathfrak{gl}(2)$ cases.

\begin{figure}[ht]
\begin{center}
\begin{pspicture}(10,4)
\rput(5,0){\rnode{B0}{$\Qf_\varnothing$}}
\rput(3,2){\rnode{B1}{$\Qf_{\{1\}}$}}
\rput(7,2){\rnode{B2}{$\Qf_{\{2\}}$}}
\rput(5,4){\rnode{B12}{$\Qf_{\{1|2\}}$}}
\ncline[ArrowInside=->,ArrowInsidePos=1.0,nodesep=0.1]{B0}{B1}
\ncline[ArrowInside=->,ArrowInsidePos=1.0,nodesep=0.1, linestyle=dashed]{B0}{B2}
\ncline[ArrowInside=->,ArrowInsidePos=1.0,nodesep=0.1, linestyle=dashed]{B1}{B12}
\ncline[ArrowInside=->,ArrowInsidePos=1.0,nodesep=0.1]{B2}{B12}
\end{pspicture}
\end{center}
\caption{Hasse diagram for the $\mathfrak{gl}(1|1)$ algebra.}
\label{Hasse11}
\end{figure}
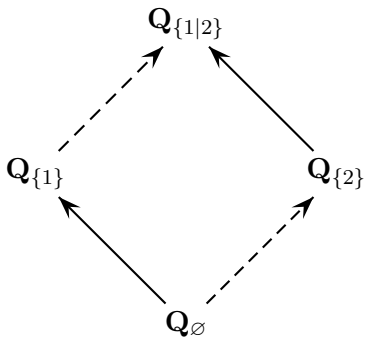

Let us first review the fusion procedure discussed in section \ref{sec:YBE} for the $\mathfrak{gl}(1|1)$ example.  Equation \eqref{fusionLL} in this case reads 
\begin{equation}
\label{fact11}
\Lbf_1(z_1)\,\Lbf_2(z_2)=\mathcal{S}\,\mathcal{L}_\varepsilon(z)\,G\,\mathcal{S}^{-1}\,,
\end{equation}
or more explicitly 
\begin{equation}
\begin{pmatrix}
z_1-\osch_1  &\,\, \oscc^{\dagger}_1  \\
-\oscc_1 &\,\, 1 \end{pmatrix} 
\begin{pmatrix}
1\,\,  & \oscc_2 \\
\oscc^{\dagger}_2\,\, & z_2+\osch_2\end{pmatrix} = 
 e^{\oscc^{\dagger}_1 \oscc^{\dagger}_2} 
\begin{pmatrix}
z+\varepsilon-\osch_1  &\,\, -2\,\varepsilon\,\oscc^{\dagger}_1\\
-\oscc_1 &\,\, z-\varepsilon -\osch_1  \end{pmatrix} 
 \begin{pmatrix}
1 \,\,\, & \oscc_2\\
0\,\,\, & 1 \end{pmatrix} e^{-\oscc^{\dagger}_1
  \oscc^{\dagger}_2}\, ,
\end{equation}
where
\begin{equation}
 \varepsilon\equiv \frac{z_1-z_2}{2}\,,\qquad z\equiv \frac{z_1+z_2}{2}\,,\qquad \osch_i=\oscc_i^{\dagger}\oscc_i-\sfrac{1}{2}\,, \qquad i=1,2\,.
\end{equation}
Here all superoscillators are of fermionic type and we denoted them by $(\oscc^\dagger,\oscc)$.
This formula is the $\alg{gl}(1|1)$ analog of equation $(3.48)$ of \cite{Bazhanov:2010ts}.
Following the same construction as in \cite{Bazhanov:2010ts}, one easily finds
\begin{equation}
\label{TQQ11}
\Top^+_\varepsilon(z)=2i\,\sin\left(\frac{\Phi_1-\Phi_2}{2}\right)\,\Qop_{1}(z_1)\,\Qop_{2}(z_2)\,.
\end{equation}
It is worth stressing that the sine factor appears on the opposite side of the equation as compared to the $\alg{gl}(2)$ case. This fact
is a direct consequence of 
\begin{equation}
 \mbox{Str}\, e^{-i\phi\,\osch}=2i\,\sin \frac{\phi}{2}\,, \qquad \mbox{Tr}\, e^{-i\phi\,\osch}=\left(2i\,\sin \frac{\phi}{2}\right)^{-1}\,.
\end{equation}
To derive the needed functional relations it is important to connect the $\Qop$-operators with the \emph{known} $\Top$, namely
\begin{equation}
 \Top^{\text{singlet}}(z)=e^{i(\Phi_1-\Phi_2)z}\,z^L\,.
\end{equation}
Every $\Top^+_\varepsilon(z)$ is constructed as a supertrace over a two dimensional representation of $\alg{gl}(1|1)$ labeled by the central charge $\varepsilon$.
Let us review how the singlet (atypical) representation emerges in this case (see e.g. \cite{Schomerus:2005bf}). If the central charge vanishes, i.e.~$\varepsilon=0$, the Fock vacuum is a  one-dimensional invariant subspace of the 2-dimensional fermionic Fock space. For $\varepsilon=0$ the $\alg{gl}(1|1)$ generators then act triangularly in the Fock space\footnote{The two-dimensional representation is indecomposable but not irreducible.}. Therefore the supertrace splits into two \emph{disjoint} contributions. This implies, that 
\begin{equation}
\label{split11}
 \alg{gl}(1|1):\qquad \Top^+_{\varepsilon=0}(z)=\Top^{\text{singlet}}(z+\half)-\Top^{\text{singlet}}(z-\half)\,,
\end{equation}
where the minus sign comes from the supertrace. It is instructive to compare \eqref{split11} with its $\alg{gl}(2)$ analog
\begin{equation}
\label{split2}
 \alg{gl}(2):\qquad \Top^+_j(z)=\Top_j(z)+\Top^+_{-j-1}(z)\,, \qquad 2j \in{\mathbb Z}_{\ge0}\,.
\end{equation}
Equations \eqref{TQQ11}, \eqref{split11} immediately imply
\begin{equation}
 2i\,\sin\left(\frac{\Phi_1-\Phi_2}{2}\right)\,\Qop_{\{ 1\}}(z)\,\Qop_{\{2 \}}(z)=\Qop_{\{ 1|2\}}(z+\half)\,\Qop_{\varnothing}(z-\half)-\Qop_{\{ 1|2\}}(z-\half)\,\Qop_{\varnothing}(z+\half)\,,
\end{equation}
where
\begin{equation}
 \Qop_{\varnothing}(z)\equiv 1\,,\qquad \quad \Qop_{\{ 1|2\}}(z)\equiv \Top^{\text{singlet}}(z)\,.
\end{equation}
This relation is of a type different from the one we we had obtained earlier in the $\mathfrak{gl}(2)$ case. Interestingly, it can nevertheless still be depicted in an analogous way with the help of a Hasse diagram. We now get the diagram in Fig. \ref{Hasse11}, where the dashed lines mean that we add a fermionic index, while the solid lines are reserved for bosonic indices.


\section{Functional Relations}
\label{sec:functional}

In this section we derive functional relations for the $\Qop$-operators using results from section \ref{sec:fusion}.
The main result of the present section are equations \eqref{placket1a}, \eqref{placket1b} (see \cite{Tsuboi:2009ud}). 

Using the results from section \ref{sec:factorization} and following the same reasoning as in \cite{Bazhanov:2010jq}
one can easily derive
\begin{equation}
\label{XQQQ}
\Delta_I(\Phi)\,\Xbf^+(z,\Lambda_I) = \Qop_{A_1}(z+\lambda_{1}')\, \Qop_{A_2}(z+\lambda_{2}') \dots \Qop_{A_{|I|}}(z+\lambda_{|I|}')\,,
\end{equation}
where $\Delta_{I}(\Phi)$ is the usual super-Vandermonde determinant
 \begin{equation}
\Delta_{I}(\Phi)=\frac{\prod_{i<j\in B(I)}\left(2i\sin\left(\frac{\Phi_{A_i}-\Phi_{A_j}}{2}\right)\right)\prod_{i<j\in F(I)}\left(2i\sin\left(\frac{\Phi_{A_i}-\Phi_{A_j}}{2}\right)\right)}{\prod_{i\in B(I),j\in F(I)}\left(2i\sin\left(\frac{\Phi_{A_i}-\Phi_{A_j}}{2}\right)\right)}\, ,
\end{equation}
%
and $\Xbf^+$ has been defined in \eqref{Xplus}. 
A beautiful feature of equation \eqref{XQQQ} is that the spectral parameter shifts of each $\Qop$-operator  become representation labels, generically complex, of the $\alg{gl}(I)$ superalgebra.  
The $\alg{gl}(I)$ representations, denoted by the symbol $+$ and the label $\Lambda_I$, are of highest weight type.
They are fully determined by the existence of an highest weight state, together with
\begin{equation}
\label{hws}
 E_{AA}\,|h.w.s\rangle=\left(-1\right)^{p(A)}\,\lambda_A\,|h.w.s\rangle\,, \qquad A \in I\,,
\end{equation}
and in conjunction with the $\alg{gl}(I)$ commutation relations.
The shifted weights  $\lambda'$ in \eqref{XQQQ} are related to the weights appearing in \eqref{hws} via
\begin{equation}
 \lambda_A'\equiv \lambda_A+\rho_A\,, \qquad \rho_A\equiv\frac{1}{2}\left(\sum_{B=A+1}^{|I|}\left(-1\right)^{p(B)}-\sum_{B=1}^{A-1}\left(-1\right)^{p(B)} \right)\,.
\end{equation}

Any functional relation among $\Xbf$-operators \eqref{Z-def} could be in principle derived using \eqref{XQQQ} together with an entirely representation-theoretical analysis relating  the operators $\Xbf^{\mathfrak{Rep}}$ for a given $\alg{gl}(I)$ representation $\mathfrak{Rep}$ to the $\Xbf^+$-operators. See the discussion in section \ref{sec:Qconstruction}. However, in this paper we will follow a more direct path in order to derive a very important set of functional relation (see \eqref{placket1a}, \eqref{placket1b}), namely the so-called Q-Q relations.
Our derivation will use results from section  \ref{sec:fusion}, and the rather simple structure of $\alg{gl}(2)$ and $\alg{gl}(1|1)$ Verma modules reviewed in section \ref{sec:oneslashone}.

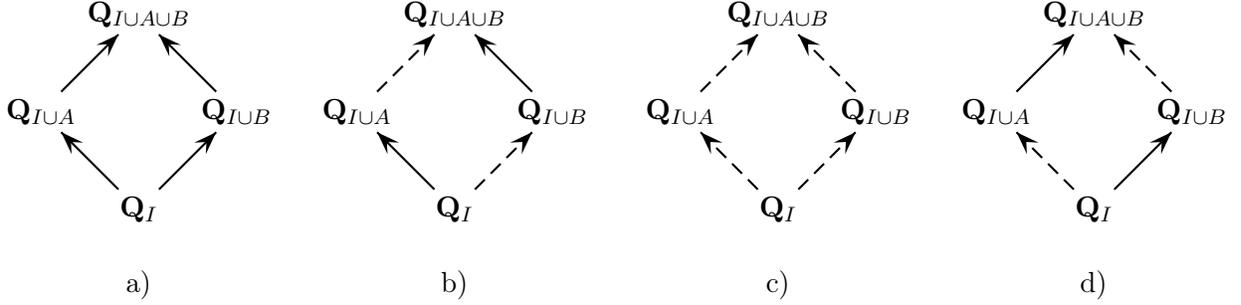
\begin{figure}[ht]
\begin{center}
\begin{pspicture}(10,4.5)
\rput(-1.2,1){\rnode{B0}{$\Qf_I$}}
\rput(-2.5,2.3){\rnode{B1}{$\Qf_{I\cup A}$}}
\rput(0.1,2.3){\rnode{B2}{$\Qf_{I\cup B}$}}
\rput(-1.2,3.6){\rnode{B12}{$\Qf_{I\cup A\cup B}$}}
\ncline[ArrowInside=->,ArrowInsidePos=1.0,nodesep=0.1]{B0}{B1}
\ncline[ArrowInside=->,ArrowInsidePos=1.0,nodesep=0.1]{B0}{B2}
\ncline[ArrowInside=->,ArrowInsidePos=1.0,nodesep=0.1]{B1}{B12}
\ncline[ArrowInside=->,ArrowInsidePos=1.0,nodesep=0.1]{B2}{B12}
\rput(3,1){\rnode{B0}{$\Qf_I$}}
\rput(1.7,2.3){\rnode{B1}{$\Qf_{I\cup A}$}}
\rput(4.3,2.3){\rnode{B2}{$\Qf_{I\cup B}$}}
\rput(3,3.6){\rnode{B12}{$\Qf_{I\cup A\cup B}$}}
\ncline[ArrowInside=->,ArrowInsidePos=1.0,nodesep=0.1]{B0}{B1}
\ncline[ArrowInside=->,ArrowInsidePos=1.0,nodesep=0.1, linestyle=dashed]{B0}{B2}
\ncline[ArrowInside=->,ArrowInsidePos=1.0,nodesep=0.1, linestyle=dashed]{B1}{B12}
\ncline[ArrowInside=->,ArrowInsidePos=1.0,nodesep=0.1]{B2}{B12}
\rput(7.3,1){\rnode{B0}{$\Qf_I$}}
\rput(6,2.3){\rnode{B1}{$\Qf_{I\cup A}$}}
\rput(8.6,2.3){\rnode{B2}{$\Qf_{I\cup B}$}}
\rput(7.3,3.6){\rnode{B12}{$\Qf_{I\cup A\cup B}$}}
\ncline[ArrowInside=->,ArrowInsidePos=1.0,nodesep=0.1, linestyle=dashed]{B0}{B1}
\ncline[ArrowInside=->,ArrowInsidePos=1.0,nodesep=0.1, linestyle=dashed]{B0}{B2}
\ncline[ArrowInside=->,ArrowInsidePos=1.0,nodesep=0.1, linestyle=dashed]{B1}{B12}
\ncline[ArrowInside=->,ArrowInsidePos=1.0,nodesep=0.1, linestyle=dashed]{B2}{B12}
\rput(11.5,1){\rnode{B0}{$\Qf_I$}}
\rput(10.2,2.3){\rnode{B1}{$\Qf_{I\cup A}$}}
\rput(12.8,2.3){\rnode{B2}{$\Qf_{I\cup B}$}}
\rput(11.5,3.6){\rnode{B12}{$\Qf_{I\cup A\cup B}$}}
\ncline[ArrowInside=->,ArrowInsidePos=1.0,nodesep=0.1, linestyle=dashed]{B0}{B1}
\ncline[ArrowInside=->,ArrowInsidePos=1.0,nodesep=0.1]{B0}{B2}
\ncline[ArrowInside=->,ArrowInsidePos=1.0,nodesep=0.1]{B1}{B12}
\ncline[ArrowInside=->,ArrowInsidePos=1.0,nodesep=0.1, linestyle=dashed]{B2}{B12}
\rput(-1.2,0){\rnode{AA}{a)}}
\rput(3,0){\rnode{BB}{b)}}
\rput(7.3,0){\rnode{CC}{c)}}
\rput(11.5,0){\rnode{DD}{d)}}
\end{pspicture}
\end{center}
\caption{Different types of the Hasse plackets: a) bosonic-bosonic, b) bosonic-fermionic, c) fermionic-fermionic, d) fermionic-bosonic.}
\label{Hasseplackets}
\end{figure}

Let us focus on equations  \eqref{fusion1}, \eqref{fusion2}. The left hand side of  \eqref{fusion1}, \eqref{fusion2} gives, upon taking the appropriate regulated trace according to section  \ref{sec:Qconstruction},
\begin{equation}
 (i):\,\,\,\Qop_{I\cup A}(z+\omega_1)\,\Qop_{I\cup B}(z+\omega_2)\,, \qquad (ii):\,\,\, \Qop_{I\cup A\cup B}(z+\omega'_1)\,\Qop_{I}(z+\omega'_1)\,,
\end{equation}
respectively.
What about the right hand side of the same equation?
As stressed in section  \ref{sec:fusion}, if $\omega_1+\omega_2=\omega'_1+\omega'_2$, the right hand sides of  \eqref{fusion1}, \eqref{fusion2} differ \emph{only} by the way the $\alg{gl}(A\cup \,B)$ algebra is realized and a \emph{decoupled} $G$-matrix.
When taking traces, equations \eqref{fusion1}, \eqref{fusion2} respectively give the following structure of auxiliary spaces
\begin{equation}
\label{trace1}
 \nstr_{\mathcal{F}^\star}\,\,\mbox{Str}_{\pi^+_\Lambda}\,\,\nstr_{\text{osc in $G$}}\,,
\end{equation}
\begin{equation}
\label{trace2}
 \nstr_{\mathcal{F}^\star}\,,
\end{equation}
where $\mathcal{F}^{\star}$ is the same in the two cases.
The relation between \eqref{trace1} and \eqref{trace2} neatly reduces to the relation between the representation $\pi^{+}_{\Lambda}$ and the singlet representation of $\alg{gl}(A\cup \, B)$ in  \eqref{gl2} and \eqref{gl11}. This point has been analyzed in some details in section \ref{sec:oneslashone} for the two rather different basic cases, namely $\alg{gl}(2)$ and $\alg{gl}(1|1)$.
The existence of a one-dimensional submodule invariant under the action of $\alg{gl}(\{A,B\})$ generators $(i)$, which in that case is just the Fock vacuum (see \eqref{gl2} and \eqref{gl11}) implies
\begin{align}
& p(A)=p(B)\,,\qquad \qquad \qquad \alg{gl}(2): \qquad \omega_1-\omega_2=\left(-1\right)^{p(A)}\,,\\
& p(A)\neq p(B)\,, \qquad \qquad \qquad \alg{gl}(1|1): \qquad \omega_1-\omega_2 =0\,.
\end{align}
This condition, together with the requirement that the action of the  $\alg{gl}(\{A,B\})$ generators  $(i)$ on this one-dimensional submodule should be the same as the one of the generators $(ii)$ in  \eqref{gl2} and \eqref{gl11} entirely
fixes\footnote{They are fixed  up to an overall shift that can be reabsorbed in the definition of the spectral parameter $z$.} the shifts $\omega_1$,  $\omega_2$,  $\omega'_1$,  $\omega'_2$.
The subtraction of Verma module is then precisely the same as in \eqref{split2} and \eqref{split11}.
Upon carefully dealing with the $\Dbf_I$ factors (see \eqref{D-exp}) and keeping track of normalizations one immediately obtains the Q-Q relations written below.
As nicely depicted in  Fig. \ref{Hasseplackets} four different cases as to be considered separately, namely
\begin{equation}
 \left( p(A),p(B)\right)\in \{(0,0);\,(1,1);\,(1,0);\,(0,1)\}\,,
\end{equation}
corresponding to four different types of Hasse plaquettes.

\noindent
These four diagrams correspond to two distinct types of relations: bosonic-bosonic or fermionic-fermionic type
\begin{equation}\label{placket1a}
(-1)^A 2i\sin\left(\frac{\Phi_A-\Phi_B}{2}\right)\,\Qop_{I\cup A\cup B}(z)\,\Qop_{I}(z)=\Qop_{I\cup A}(z+\sfrac{1}{2})\,\Qop_{I\cup B}(z-\sfrac{1}{2})-\Qop_{I\cup A}(z-\sfrac{1}{2})\,\Qop_{I\cup B}(z+\sfrac{1}{2})
\end{equation}
and bosonic-fermionic or fermionic-bosonic type
\begin{equation}\label{placket1b}
(-1)^A 2i\sin\left(\frac{\Phi_A-\Phi_B}{2}\right)\,\Qop_{I\cup A}(z)\,\Qop_{I\cup B}(z)=\Qop_{I\cup A\cup B}(z+\sfrac{1}{2})\,\Qop_{I}(z-\sfrac{1}{2})-\Qop_{I\cup A \cup B}(z-\sfrac{1}{2})\,\Qop_{I}(z+\sfrac{1}{2})
\end{equation}
This knowledge is enough to draw the Hasse diagram for any algebra $\mathfrak{gl}(n|m)$. As we will show in the next section, the relations \eqref{placket1a} and \eqref{placket1b} are sufficient to derive the Bethe equations.

We would like to add here an intriguing observation. One immediately notices that the relations \eqref{placket1a} and \eqref{placket1b} look exactly the same if we rename the indices of the $\Qop$-operators. In fact, relation \eqref{placket1a} tells us that when taking a product of the upper $\Qop$-operator with the lower one in Fig.~\ref{Hasseplackets}a, this then equals the difference of products of the right and left $\Qop$-operators with appropriately shifted arguments. The formula \eqref{placket1b} gives similar information about figure \ref{Hasseplackets}b if we formally exchange the upper and lower with the right side and left side $\Qop$-operators. On the other hand, if we relabel the $\Qop$-operators in Fig.~\ref{Hasseplackets}a, the relation stemming from \eqref{placket1b} will have the same interpretation as the one of \eqref{placket1a}. This can be seen when clockwise turning the Hasse plaquette in Fig.~\ref{Hasseplackets}b by $90^\circ$. This rotation changes fermionic lines to bosonic ones, and inverts the arrows. Analogously, turning the plaquette in Fig.~\ref{Hasseplackets}c by $180^\circ$, we obtain once again the bosonic-like Hasse plaquette described by the same relation \eqref{placket1a}. This way we can rotate the entire Hasse diagram such that all lines will be bosonic, and we will end up with the situation known from the $\mathfrak{gl}(n)$ Hasse diagram. It is known that such Hasse diagrams can be solved by determinant formulas. The only difference is that the determinants we will get here will be built from non-partonic objects, as it was the case for $\mathfrak{gl}(n)$. We will aptly call the procedure presented above ``bosonization of the Hasse diagram'' (compare with \cite{Gromov:2010km}). We can also fermionize a Hasse diagram by rotating it such that all lines will be fermionic. This case leads to determinant formulas as well.


\section{Bethe Equations}
\label{sec:bethe}

The derivation of the Bethe equations of the supersymmetric $\mathfrak{gl}(n|m)$ spin chains from the hierarchy of Baxter operators proceeds in much the same way as in the $\mathfrak{gl}(n)$ case \cite{Bazhanov:2010jq}. Once again it is very useful to work with hypercubic Hasse diagrams. We simply have to consider any path on the Hasse diagram leading from $\Qop_{\{1,\ldots,n+m\}}$ to $\Qop_\varnothing$. Each such path corresponds to a set of nested  Bethe equations. In total there are $(n+m)!$ different paths and thus the same number of sets of equations. There is one major difference in comparison with $\mathfrak{gl}(n)$ case. In the latter all sets of the Bethe equations look the same, in line with the fact that there is a unique Dynkin diagram for the $\mathfrak{gl}(n)$ algebra. In the supersymmetric case we have distinct Dynkin diagrams which differ by the various possible gradings of the diagram nodes. A white node of the diagram corresponds to a doublet (one positive and one negative root) of bosonic simple roots, and a crossed node to a doublet of fermionic simple roots of the $\mathfrak{gl}(n|m)$ algebra. Clearly there is a minimum of one crossed node, while the other extreme is that all nodes are fermionic. In the Hasse diagram picture these differences are encoded in the order of dashed and solid lines along the chosen path. 

Now there are two distinct types of equations we get when taking ratios of the Q-Q relations, which are associated to the plaquettes of the hypercubic Hasse diagram, at special points of the spectral parameter $z$. For the Q-Q relations of type \eqref{placket1a} we get the same equation as in the $\mathfrak{gl}(n)$ case
\beq
\label{formalBETHE}
-1=\frac{{\rm Q}_{I }(\specbaz_k^{I \cup A}-\half)}{{\rm Q}_{I }(\specbaz_k^{I \cup A}+\half)}\,
\frac{ {\rm Q}_{I \cup A}(\specbaz_k^{I \cup A}+1)}{ {\rm Q}_{I \cup A}(\specbaz_k^{I \cup A}-1)}\,
\frac{{\rm Q}_{I\cup A \cup B}(\specbaz_k^{I \cup A}-\half)}{{\rm Q}_{I\cup A \cup B}(\specbaz_k^{I \cup A}+\half)}\,.
\eeq
On the other hand, for the Q-Q relations \eqref{placket1b}, when evaluating at  $z=z_k^{I\cup A}$, we get
\begin{equation}\label{formalBETHE2}
1=\frac{Q_{I}(z_k^{I\cup A}+\sfrac{1}{2})\,Q_{I\cup A\cup B}(z_k^{I\cup A}-\sfrac{1}{2})}{Q_{I}(z_k^{I\cup A}-\sfrac{1}{2})\,Q_{I\cup A\cup B}(z_k^{I\cup A}+\sfrac{1}{2})}\, .
\end{equation}
For a given path we get a, in general, mixed set of equations of both types \eqref{formalBETHE} and \eqref{formalBETHE2}, depending on which path we take. For any node on the path we have to take a look at the Hasse diagram and check if the incoming and outgoing lines on the path are of the same, or a different type. In the former situation we write equation \eqref{formalBETHE},  and in the latter equation \eqref{formalBETHE2}.  This way we can immediately read off all possible sets of Bethe equations from the Hasse diagram. It is important to stress that all these $(n+m)!$ sets, despite the fact that they will look rather different, will give {\it exactly} the same solution of our spectral problem.

Let us now rewrite the Bethe equations in their traditional form. It follows from our construction that the $\Qop$-operator is a polynomial in the spectral parameter $z$, with some exponent normalization factor
\begin{equation}
\label{Q-eigen}
{\rm Q}_I(z)=e^{i z\,(\,\sum_{A\in I}(-1)^{A}\Phi_A)} \, \prod_k
(z-z^I_k)\, .
\end{equation} 
We would like to stress once more that this absolutely crucial statement on the analytic structure of the eigenvalues of the $\Qop$-operator is not {\it assumed}, but obtained by construction! Plugging this into the relations \eqref{formalBETHE} and \eqref{formalBETHE2}, we will get
\begin{equation}
e^{(-1)^{A_{i+1}}i\,(\Phi_{A_{i+1}}-\Phi_{A_i})}=\prod_k\frac{z_l^{I_i}-z_k^{I_{i-1}}-\sfrac{1}{2}}{z_l^{I_i}-z_k^{I_{i-1}}+\sfrac{1}{2}}\prod_{k\neq l}\frac{z_l^{I_i}-z_k^{I_i}+1}{z_l^{I_i}-z_k^{I_i}-1}\prod_k\frac{z_l^{I_i}-z_k^{I_{i+1}}-\sfrac{1}{2}}{z_l^{I_i}-z_k^{I_{i+1}}+\sfrac{1}{2}}
\end{equation}
for the bosonic-bosonic or fermionic-fermionic node on the Hasse diagram, and
\begin{equation}
e^{(-1)^{A_{i+1}}i\,(\Phi_{A_{i+1}}-\Phi_{A_i})}=\prod_k\frac{z_l^{I_i}-z_k^{I_{i-1}}+\sfrac{1}{2}}{z_l^{I_i}-z_k^{I_{i-1}}-\sfrac{1}{2}}\prod_k\frac{z_l^{I_i}-z_k^{I_{i+1}}-\sfrac{1}{2}}{z_l^{I_i}-z_k^{I_{i+1}}+\sfrac{1}{2}}
\end{equation}
for the bosonic-fermionic or fermionic-bosonic node on the Hasse diagram. 
The Bethe equations corresponding to the lowest and highest level of the nested system can be obtained using the a priori knowledge of the $\Qop$-operators at the ``top'' and ``bottom'' of the Hasse diagram:
\begin{equation}
 \Qop_{\varnothing}=1\,, \qquad \Qop_{\{1,\dots,n+m\}}=e^{iz\,\sum_A(-1)^A\,\Phi_A}\,z^L\,.
\end{equation}

To conclude our solution procedure for the $\alg{gl}(n|m)$-spin chain we just state the well-known
expression for the eigenvalues of \eqref{slnm-ham} (or equivalently \eqref{slnm-ham-alt} of the Hamiltonian of the graded spin chain. It only involves the roots $z^{I_{n+m-1}}$ of any of the $n+m$ possible sets $I_{n+m-1}$ on the last-level of the nested Bethe equations:
\begin{equation}\label{energy.formula}
E_{n|m}=2\sum_{k=1}^{m_{n+m-1}}\frac{1}{\frac{1}{4}-\left(z_{k}^{I_{n+m-1}}\right)^2}\, ,
\quad {\rm or} \quad
E_{n|m}=4\,L-2\sum_{k=1}^{m_{n+m-1}}\frac{1}{\frac{1}{4}-\left(z_{k}^{I_{n+m-1}}\right)^2}\, .
\end{equation}
Here $m_{n+m-1}$ is the number of roots of  the $Q_{I_{n+m-1}}(z)$ function. The left expression in \eqref{energy.formula} is for a bosonic vacuum, c.f.~\eqref{bose-fermi-vac}, which corresponds to the case where $I_{n+m-1}$ is such that one of the first $n$ ``bosonic'' indices is missing from the set $\{ 1,2,\ldots ,n+m\}$. The right expression in \eqref{energy.formula} is, in view of the non-trivial vacuum energy of the r.h.s.~of \eqref{bose-fermi-vac}, for a fermionic vacuum, which corresponds to the case where $I_{n+m-1}$ is such that one of the $m$ ``fermionic'' indices $\{ n+1,n+2,\ldots ,n+m\}$ is missing from the set $\{ 1,2,\ldots ,n+m\}$.


\section{ \texorpdfstring{$\mathfrak{gl}(2|1)$}{}}
\label{sec:tJ}

To illustrate some of the content of the previous sections we will present here the application of our formalism to the case of the $\mathfrak{gl}(2|1)$ algebra. In this case both the $\mathfrak{gl}(2)$ and $\mathfrak{gl}(1|1)$ Q-Q relations appear in the analysis. Physically it corresponds to the diagonalization of the t-J model, which was first solved by Bethe ansatz in \cite{Schlottmann:1987zz},\cite{Essler:1992he, Foerster:1992uk}. A major part of this section can also be found in other papers, see e.g. \cite{Kazakov:2007fy, Bazhanov:2008yc}.  

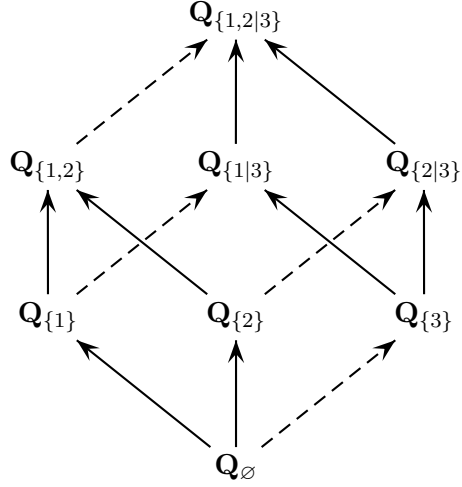
\begin{figure}[t]
\begin{center}
\begin{pspicture}(8,6)
\rput(4,0){\rnode{A0}{$\Qf_{\varnothing}$}}
\rput(1.5,2){\rnode{A1}{$\Qf_{\{1\}}$}}
\rput(4,2){\rnode{A2}{$\Qf_{\{2\}}$}}
\rput(6.5,2){\rnode{A3}{$\Qf_{\{3\}}$}}
\rput(1.5,4){\rnode{A12}{$\Qf_{\{1,2\}}$}}
\rput(4,4){\rnode{A13}{$\Qf_{\{1|3\}}$}}
\rput(6.5,4){\rnode{A23}{$\Qf_{\{2|3\}}$}}
\rput(4,6){\rnode{A123}{$\Qf_{\{ 1,2|3\}}$}}
\ncline[ArrowInside=->,ArrowInsidePos=1.0,nodesep=0.1]{A0}{A1}
\ncline[ArrowInside=->,ArrowInsidePos=1.0,nodesep=0.1]{A0}{A2}
\ncline[ArrowInside=->,ArrowInsidePos=1.0,nodesep=0.1, linestyle=dashed]{A0}{A3}
\ncline[ArrowInside=->,ArrowInsidePos=1.0,nodesep=0.1]{A1}{A12}
\ncline[ArrowInside=->,ArrowInsidePos=1.0,nodesep=0.1, linestyle=dashed]{A1}{A13}
\ncline[ArrowInside=->,ArrowInsidePos=1.0,nodesep=0.1]{A2}{A12}
\ncline[ArrowInside=->,ArrowInsidePos=1.0,nodesep=0.1, linestyle=dashed]{A2}{A23}
\ncline[ArrowInside=->,ArrowInsidePos=1.0,nodesep=0.1]{A3}{A13}
\ncline[ArrowInside=->,ArrowInsidePos=1.0,nodesep=0.1]{A3}{A23}
\ncline[ArrowInside=->,ArrowInsidePos=1.0,nodesep=0.1, linestyle=dashed]{A12}{A123}
\ncline[ArrowInside=->,ArrowInsidePos=1.0,nodesep=0.1]{A13}{A123}
\ncline[ArrowInside=->,ArrowInsidePos=1.0,nodesep=0.1]{A23}{A123}
\end{pspicture}
\caption{Hasse diagram for the $\mathfrak{gl}(2|1)$ algebra.}
\label{Hasse21}
\end{center}
\end{figure}

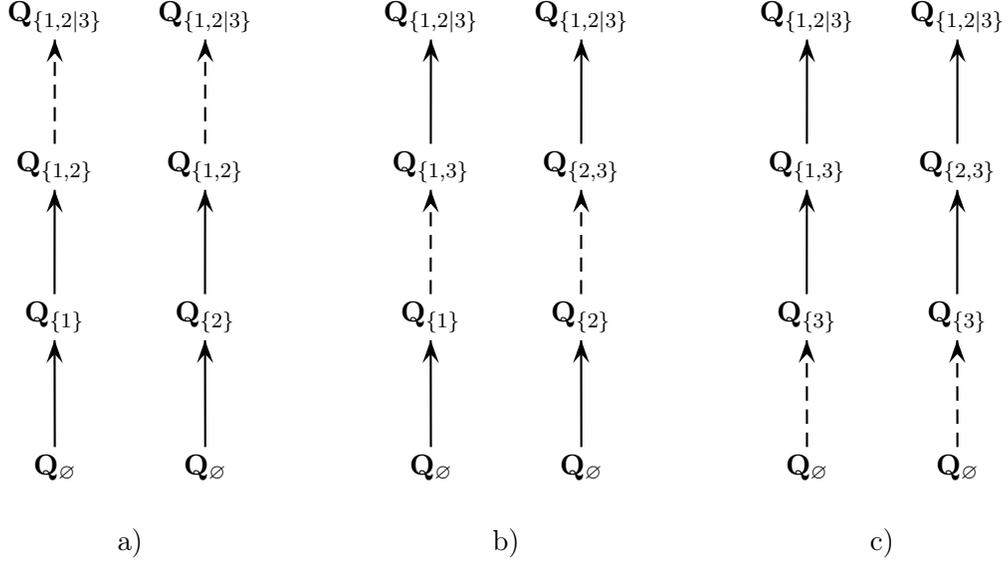
\begin{figure}[t]
\begin{center}
\begin{pspicture}(12,7)
\rput(0,1){\rnode{A0}{$\Qf_{\varnothing}$}}
\rput(0,3){\rnode{A1}{$\Qf_{\{1\}}$}}
\rput(0,5){\rnode{A12}{$\Qf_{\{1,2\}}$}}
\rput(0,7){\rnode{A123}{$\Qf_{\{ 1,2|3\}}$}}
\ncline[ArrowInside=->,ArrowInsidePos=1.0,nodesep=0.1]{A0}{A1}
\ncline[ArrowInside=->,ArrowInsidePos=1.0,nodesep=0.1]{A1}{A12}
\ncline[ArrowInside=->,ArrowInsidePos=1.0,nodesep=0.1, linestyle=dashed]{A12}{A123}
\rput(2,1){\rnode{B0}{$\Qf_{\varnothing}$}}
\rput(2,3){\rnode{B1}{$\Qf_{\{2\}}$}}
\rput(2,5){\rnode{B12}{$\Qf_{\{1,2\}}$}}
\rput(2,7){\rnode{B123}{$\Qf_{\{ 1,2|3\}}$}}
\ncline[ArrowInside=->,ArrowInsidePos=1.0,nodesep=0.1]{B0}{B1}
\ncline[ArrowInside=->,ArrowInsidePos=1.0,nodesep=0.1]{B1}{B12}
\ncline[ArrowInside=->,ArrowInsidePos=1.0,nodesep=0.1, linestyle=dashed]{B12}{B123}
\rput(5,1){\rnode{C0}{$\Qf_{\varnothing}$}}
\rput(5,3){\rnode{C1}{$\Qf_{\{1\}}$}}
\rput(5,5){\rnode{C12}{$\Qf_{\{1,3\}}$}}
\rput(5,7){\rnode{C123}{$\Qf_{\{ 1,2|3\}}$}}
\ncline[ArrowInside=->,ArrowInsidePos=1.0,nodesep=0.1]{C0}{C1}
\ncline[ArrowInside=->,ArrowInsidePos=1.0,nodesep=0.1, linestyle=dashed]{C1}{C12}
\ncline[ArrowInside=->,ArrowInsidePos=1.0,nodesep=0.1]{C12}{C123}
\rput(7,1){\rnode{D0}{$\Qf_{\varnothing}$}}
\rput(7,3){\rnode{D1}{$\Qf_{\{2\}}$}}
\rput(7,5){\rnode{D12}{$\Qf_{\{2,3\}}$}}
\rput(7,7){\rnode{D123}{$\Qf_{\{ 1,2|3\}}$}}
\ncline[ArrowInside=->,ArrowInsidePos=1.0,nodesep=0.1]{D0}{D1}
\ncline[ArrowInside=->,ArrowInsidePos=1.0,nodesep=0.1, linestyle=dashed]{D1}{D12}
\ncline[ArrowInside=->,ArrowInsidePos=1.0,nodesep=0.1]{D12}{D123}
\rput(10,1){\rnode{E0}{$\Qf_{\varnothing}$}}
\rput(10,3){\rnode{E1}{$\Qf_{\{3\}}$}}
\rput(10,5){\rnode{E12}{$\Qf_{\{1,3\}}$}}
\rput(10,7){\rnode{E123}{$\Qf_{\{ 1,2|3\}}$}}
\ncline[ArrowInside=->,ArrowInsidePos=1.0,nodesep=0.1, linestyle=dashed]{E0}{E1}
\ncline[ArrowInside=->,ArrowInsidePos=1.0,nodesep=0.1]{E1}{E12}
\ncline[ArrowInside=->,ArrowInsidePos=1.0,nodesep=0.1]{E12}{E123}
\rput(12,1){\rnode{F0}{$\Qf_{\varnothing}$}}
\rput(12,3){\rnode{F1}{$\Qf_{\{3\}}$}}
\rput(12,5){\rnode{F12}{$\Qf_{\{2,3\}}$}}
\rput(12,7){\rnode{F123}{$\Qf_{\{ 1,2|3\}}$}}
\ncline[ArrowInside=->,ArrowInsidePos=1.0,nodesep=0.1, linestyle=dashed]{F0}{F1}
\ncline[ArrowInside=->,ArrowInsidePos=1.0,nodesep=0.1]{F1}{F12}
\ncline[ArrowInside=->,ArrowInsidePos=1.0,nodesep=0.1]{F12}{F123}
\rput(1,0){\rnode{AA}{a)}}
\rput(6,0){\rnode{BB}{b)}}
\rput(11,0){\rnode{CC}{c)}}
\end{pspicture}
\end{center}
\caption{All different paths in the Hasse diagram of $\mathfrak{gl}(2|1)$ algebra.}
\label{fig:paths}
\end{figure}

For the $\mathfrak{gl}(2|1)$ algebra we deal with two bosonic indices and one fermionic index. There are 8 different $\Qop$-operators 
\begin{equation}
\Qop_{\varnothing},\,\Qop_{\{1\}},\,\Qop_{\{2\}},\,\Qop_{\{3\}},\,\Qop_{\{1,2\}},\,\Qop_{\{1|3\}},\,\Qop_{\{2|3\}},\,\Qop_{\{1,2|3\}}\, ,
\end{equation}
which form the cubic Hasse diagram depicted in Figure \ref{Hasse21}\,.
In order to derive Bethe equations for the $\mathfrak{gl}(2|1)$ algebra we will be interested in paths starting from $\Qop_{\varnothing}$ and leading to $\Qop_{\{ 1,2|3\}}$. There are six such paths on the Hasse diagram, while there are three different Dynkin diagrams of $\mathfrak{gl}(2|1)$. Each Dynkin diagram corresponds to two paths, related by the  $\mathfrak{gl}(2)$ symmetry between the bosonic indices 1 and 2. The paths are presented in the Figure \ref{fig:paths}. 

Let us present here the three different sets of Bethe equations. We see that the various types of Bethe equations correspond to the different distributions of the fermionic nodes on the Dynkin diagram\footnote{We use the standard notation where an empty node is bosonic and a crossed node is fermionic.}. In the language of the Hasse diagrams it corresponds to the different orders of the bosonic and fermionic lines on the paths in Fig. \ref{fig:paths}. We mark the momentum carrying node by putting 1 next to it.

\begin{minipage}{0.1\linewidth}\centering
\begin{picture}(15,55)
\put(-15,45){1}
\put(0,0){\circle{20}}
\put(0,40){\circle{20}}
\put(0,10){\line(0,1){20}}
\put(-7.07,32.93){\line(1,1){14.14}}
\put(7.07,32.93){\line(-1,1){14.14}}
\end{picture}
\end{minipage}
\begin{minipage}{0.85\linewidth}
\begin{eqnarray}
\frac{Q_{\{ 1,2|3\}}\left(z^{\{ 1,2\}}_k+\sfrac{1}{2}\right)}{Q_{\{1,2|3\}}\left(z^{\{ 1,2\}}_k-\sfrac{1}{2}\right)}&=&\frac{Q_{\{ 1\}}\left(z^{\{ 1,2\}}_k+\sfrac{1}{2}\right)}{Q_{\{ 1\}}\left(z^{\{ 1,2\}}_k-\sfrac{1}{2}\right)}\\
-1&=&\frac{Q_{\{ 1,2\}}\left(z^{\{ 1\}}_k-\sfrac{1}{2}\right)}{Q_{\{ 1,2\}}\left(z^{\{ 1\}}_k+\sfrac{1}{2}\right)}\frac{Q_{\{ 1\}}\left(z^{\{ 1\}}_k+1\right)}{Q_{\{ 1\}}\left(z^{\{ 1\}}_k-1\right)}
\end{eqnarray}
\end{minipage}

\begin{minipage}{0.1\linewidth}\centering
\begin{picture}(15,55)
\put(-15,45){1}
\put(0,0){\circle{20}}
\put(0,40){\circle{20}}
\put(0,10){\line(0,1){20}}
\put(-7.07,-7.07){\line(1,1){14.14}}
\put(7.07,-7.07){\line(-1,1){14.14}}
\put(-7.07,32.93){\line(1,1){14.14}}
\put(7.07,32.93){\line(-1,1){14.14}}
\end{picture}
\end{minipage}
\begin{minipage}{0.85\linewidth}
\begin{eqnarray}
\hspace{-2.9cm}\frac{Q_{\{ 1,2|3\}}\left(z^{\{ 1|3\}}_k+\sfrac{1}{2}\right)}{Q_{\{1,2|3\}}\left(z^{\{ 1|3\}}_k-\sfrac{1}{2}\right)}&=&\frac{Q_{\{ 1\}}\left(z^{\{ 1|3\}}_k+\sfrac{1}{2}\right)}{Q_{\{ 1\}}\left(z^{\{|1,3\}}_k-\sfrac{1}{2}\right)}\\
1&=&\frac{Q_{\{ 1|3\}}\left(z^{\{ 1\}}_k-\sfrac{1}{2}\right)}{Q_{\{ 1|3\}}\left(z^{\{ 1\}}_k+\sfrac{1}{2}\right)}
\end{eqnarray}
\end{minipage}

\begin{minipage}{0.1\linewidth}\centering
\begin{picture}(15,55)
\put(-15,45){1}
\put(0,0){\circle{20}}
\put(0,40){\circle{20}}
\put(0,10){\line(0,1){20}}
\put(-7.07,-7.07){\line(1,1){14.14}}
\put(7.07,-7.07){\line(-1,1){14.14}}
\end{picture}
\end{minipage}
\begin{minipage}{0.85\linewidth}
\begin{eqnarray}
\frac{Q_{\{ 1,2|3\}}\left(z^{\{ 1|3\}}_k+\sfrac{1}{2}\right)}{Q_{\{1,2|3\}}\left(z^{\{ 1|3\}}_k-\sfrac{1}{2}\right)}&=&\frac{Q_{\{ 3\}}\left(z^{\{ 1|3\}}_k-\sfrac{1}{2}\right)}{Q_{\{ 3\}}\left(z^{\{ 1|3\}}_k+\sfrac{1}{2}\right)}\frac{Q_{\{ 1|3\}}\left(z^{\{ 1|3\}}_k+1\right)}{Q_{\{ 1|3\}}\left(z^{\{ 1|3\}}_k-1\right)}\\
1&=&\frac{Q_{\{ 1|3\}}\left(z^{\{ 3\}}_k-\sfrac{1}{2}\right)}{Q_{\{1|3\}}\left(z^{\{ 3\}}_k+\sfrac{1}{2}\right)}\end{eqnarray}
\end{minipage}

A final comment about the $\mathfrak{gl}(2|1)$ algebra is that, according to the discussion from section \ref{sec:functional}, upon rotating the Hasse diagram in the Fig.~\ref{Hasse21} such that the operator $\Qop_{\{ 3\}}$ will be the base of the cube, we will get a Hasse diagram with just bosonic lines. Such a diagram can be solved in terms of determinants, which leads us to a determinant formula for all operators. These may all be written in terms of the lowest two layers of the Hasse diagram, which are given by the operators $\Qop_{\{ 3\}}, \Qop_{\{ 1|3\}},\Qop_{\{ 2|3\}}$, and $\Qop_{\varnothing}$. 


\subsection*{Acknowledgments}
We thank Vladimir Bazhanov, Volodya Kazakov, Vladimir Mitev, and Zengo Tsuboi for very useful discussions. T.~{\L}ukowski is supported by a DFG grant in the framework of the SFB 647 {\it ``Raum - Zeit - Materie. Analytische und Geometrische Strukturen''}. 

\appendix

\section{Details for Section \ref{sec:fusion}}
\label{app:details}
The $G$ matrix in \eqref{quadspec} is given by 
\begin{equation}
G=\left(
\begin{BMAT}(r)[0.15cm,0cm,0cm]{c.c.c}{c.c.c}
\delta_{A B} &0 &0\\
0&\ g_{\dot{A}\dot{B}} &0\\
0&  0& \ \delta_{\ddot{A}\ddot{B}}
\end{BMAT}\, \right)\\, 
\qquad 
g_{\dot{A}\dot{B}}=
\left(
\begin{BMAT}(r)[0.2cm,0.2cm]{c.c}{c.c}
\delta_{\dot{A}_1\dot{B}_1} & -\left(-1 \right)^{\dot{B}_2}\,\suposc^{[2]}_{\dot{A}_1 \dot{B}_2} \\
0 & \delta_{\dot{A}_2\dot{B}_2}\\
\end{BMAT}\, \right)\\,  
\end{equation}
The similarity transform $\mathcal{S}$ in \eqref{quadspec} is given by
\begin{equation}
 \mathcal{S}=\mathcal{S}_0\,\mathcal{S}_1\,\mathcal{S}_2\,\mathcal{S}_3\,,
\end{equation}
with
\begin{equation}
 \mathcal{S}_0=\exp \left[(-1)^{\dot A_1}\suposc^{\dagger[1]}_{ \dot A_1{\dot B_2}}\,\suposc^{\dagger[2]}_{{\dot B_2}\dot A_1}\right]\,,
\end{equation}
\begin{equation}
 \mathcal{S}_1= \exp \left[-(-1)^{\dot C_2}\suposc^{\dagger[2]}_{{A}\ddot C }\,\suposc^{[2]}_{{\ddot C}\dot C_2 }\,\suposc^{[1]}_{\raisemath{-4pt}{{\dot C_2}A}}-(-1)^{C+\dot C_2}\suposc^{\dagger[2]}_{{C}\dot C_1 }\,\suposc^{\dagger[1]}_{{\dot C_1}\dot C_2 }\,\suposc^{[1]}_{\raisemath{-4pt}{{\dot C_2}C}}-(-1)^{C+\ddot C}\suposc^{\dagger[2]}_{{C}\dot C_1 }\,\suposc^{\dagger[1]}_{{\dot C_1}\ddot C }\,\suposc^{[1]}_{\raisemath{-4pt}{{\ddot C}C}}\right]\,,
\end{equation}  
\begin{equation}
 \mathcal{S}_2=\exp \left[\suposc^{\dagger[1]}_{{\dot C_1}\dot C_2 }\,\suposc^{\dagger[2]}_{{\dot C_2}\ddot C }\,\left(\mathcal{I}^{-1} \right)_{\raisemath{-2pt}{\ddot{C}\ddot{A}}}\,\suposc^{[1]}_{\raisemath{-4pt}{{\ddot A}\dot C_1}}\right]\,,
\end{equation}
\begin{equation}
 \mathcal{S}_3=\exp \left[\suposc^{\dagger[1]}_{{\dot A_1}\ddot B}\,\left(\log\,\mathcal{I}\right)_{\ddot{B}\ddot{C}}\suposc^{[1]}_{{\ddot C}\dot{A}_1}\,\right]\,,
\end{equation}
\begin{equation}
 \mathcal{I}_{\ddot{A}\ddot{B}}\equiv \delta_{\ddot{A}\ddot{B}}-
(-1)^{C}\suposc^{[1]}_{\ddot A C}\suposc^{\dagger [2]}_{ C\ddot{B}}\,.
\end{equation}
Despite the rather complicated structure of the similarity transform $\mathcal{S}$, the function of its constituents is rather neat.
$\mathcal{S}_0$ is introduced in order to disentangle the oscillators in the $G$ matrix from the remaining oscillators in \eqref{quadspec}.
$\mathcal{S}_1$ is choosen to have $L^{(1)}$ in the canonical form \eqref{L1fusion}. 
$\mathcal{S}_2$ and $\mathcal{S}_3$ are introduced in order to have the 
$L^{(2)}$ elements in equations \eqref{L2easy}, \eqref{L2easybis} in that canonical form. Let us stress that $\mathcal{S}_2$ and $\mathcal{S}_3$ do not act on $L^{(1)}$.

\subsection{Identification of Oscillators}
This appendix contains the explicit identification of the superoscillators in equations \eqref{L1fusion},
\eqref{L1fusionbis}, \eqref{L2fusion}, \eqref{L2easy}, \eqref{L2easybis} with the superoscillators in \eqref{prodlax}. The identification is
\begin{equation}
 \left(\suposc_{ \dot{A}_1 B},\,\, \suposc_{ \dot{A}_2 B},\,\, \suposc_{ \ddot{A} B},\,\, \tilde{\suposc}_{ \ddot{A} B},\, \suposc_{ \ddot{A} \dot{B}_1},\, \suposc_{ \ddot{A} \dot{B}_2}\right)=
  \left(\suposc^{[2]}_{ \dot{A}_1 B},\,\, \suposc^{[1]}_{ \dot{A}_2 B},\,\, \suposc^{[2]}_{ \ddot{A} B},\,\, \suposc^{[1]}_{ \ddot{A} B},\, \suposc^{[1]}_{ \ddot{A} \dot{B}_1},\, \suposc^{[2]}_{ \ddot{A} \dot{B}_2}\right)\, .
\end{equation}
The analogous equation for  $\suposc^\dagger$ is easily obtained.


\bibliographystyle{utphys}
\bibliography{flms}

\end{document}